\documentclass[aps,onecolumn,11pt,floatfix,altaffilletter,superscriptaddress,preprintnumbers, tightenlines,showpacs,showkeys,notitlepage,nofootinbib]{revtex4-2}

\usepackage[colorlinks=true,citecolor=blue,linkcolor=blue]{hyperref}
\usepackage[normalem]{ulem}
\usepackage{amsmath,amssymb}
\usepackage{mathtools}
\usepackage{verbatim}
\usepackage{titlesec}               
\usepackage{epsfig}
\usepackage{graphicx}               
\usepackage{url}
\usepackage{color}
\usepackage{multirow}
\usepackage{floatrow}
\usepackage{placeins}
\usepackage[dvipsnames]{xcolor}
\usepackage{epstopdf}
\usepackage{siunitx}
\usepackage{fontawesome}
\usepackage{tikz}
\usepackage{tikz-feynman}
\usepackage{enumitem}
\usepackage[capitalize]{cleveref}
\usepackage{lipsum}
\usepackage{gensymb}
\usepackage{booktabs}               
\usepackage{xspace}                 
\usepackage{pifont}
\usepackage{marvosym }
\usetikzlibrary{shapes,arrows}
\usetikzlibrary{decorations.pathmorphing,decorations.markings}
\usetikzlibrary{snakes}
\usepackage{orcidlink}  
\allowdisplaybreaks

\usepackage{bbm}
\usepackage{slashed}

\makeatletter

\renewcommand{\p@subsection}{}
\makeatother

\titleformat*{\section}{\centering\bfseries\uppercase}
\titlelabel{\thetitle\quad}
\titleformat*{\paragraph}{\bfseries}
\titlespacing*{\paragraph}{0pt}{3.25ex plus 1ex minus .2ex}{1em}

\makeatletter
\def\l@subsubsection#1#2{}
\makeatother

\newsavebox{\twosubbox}

\newcommand{\newtext}[1]{\textcolor{black}{ {#1}}}

\begin{document}

\title{Collider Prospects for the Neutrino Magnetic Moment Portal}

\author{Vedran Brdar \orcidlink{0000-0001-7027-5104}}
\email{vedran.brdar@okstate.edu}
\affiliation{Department of Physics, Oklahoma State University, Stillwater, OK, 74078, USA}
\author{Ying-Ying Li \orcidlink{0000-0003-2580-1974}}
\email{liyingying@ihep.ac.cn}
\affiliation{Institute of High Energy Physics, Chinese Academy of Sciences, Beijing 100049, China}
\affiliation{Interdisciplinary Center for Theoretical Study, University of Science and Technology of China, Hefei, Anhui 230026, China}
\affiliation{Peng Huanwu Center for Fundamental Theory, Hefei, Anhui 230026, China}
\author{Samiur R. Mir \orcidlink{0000-0002-6531-2174} }
\email{samiur.mir@okstate.edu}
\affiliation{Department of Physics, Oklahoma State University, Stillwater, OK, 74078, USA}
\author{Yi-Lin Wang \orcidlink{0009-0005-8351-3198}}
\email{wangyilin@mail.ustc.edu.cn}
\affiliation{Interdisciplinary Center for Theoretical Study, University of Science and Technology of China, Hefei, Anhui 230026, China}
\affiliation{Peng Huanwu Center for Fundamental Theory, Hefei, Anhui 230026, China}

\begin{abstract}
The transition magnetic moment between active and sterile neutrinos is theoretically well-motivated scenario beyond the Standard Model, which can be probed in cosmology, astrophysics, and at terrestrial experiments. In this work, we focus on the latter by examining such an interaction at proposed lepton colliders. Specifically, in addition to revisiting LEP, we consider CEPC, FCC-ee, CLIC, and the muon collider, motivated by the potential realization of any of them. Within the effective field theory framework, we present parameter regions that can be probed, highlighting the dependence on the lepton flavor interacting with the sterile neutrino. By including several new processes with large sterile neutrino production cross sections at high-energy lepton colliders, we find that the expected sensitivity for the active-to-sterile neutrino transition magnetic moment  can reach $d_\gamma \simeq \mathcal{O}(10^{-7})$ GeV$^{-1}$. 
\end{abstract}

\maketitle

\section{Introduction}
\label{sec:intro}
\noindent
Sterile neutrinos are a well-motivated extension of the Standard Model (SM) that can explain nonzero neutrino masses via the seesaw mechanism \cite{Minkowski,GellMann:1980vs,Yanagida:1979as,Goran} and the baryon asymmetry of the Universe through leptogenesis \cite{Fukugita:1986hr,Akhmedov:1998qx}. In standard scenarios, sterile neutrinos interact with the SM only through mixing, which poses significant challenges to their discovery prospects in the viable parameter space for the low-scale seesaw \cite{Bolton:2019pcu,Atre:2009rg,deGouvea:2015euy}. However, the discovery potential improves if sterile neutrinos have additional types of interaction, with the gauge-invariant neutrino magnetic moment portal being particularly appealing. Such an interaction was comprehensively studied in the context of cosmological, astrophysical, and terrestrial probes; see \emph{e.g.} \cite{Magill:2018jla,Brdar:2020quo} and references therein. 

The constraints associated with terrestrial experiments have the smallest uncertainties, making these probes particularly robust. Among them, in the present work, we will focus on the energy frontier by studying collider prospects for the neutrino magnetic moment portal. Our primary focus will be on the proposed lepton colliders: CEPC, FCC-ee, CLIC, and the muon collider (MuC). We will examine a number of sterile neutrino production channels at these experiments, focusing on the parameter space where sterile neutrinos decay promptly, within the central detector. We will present expected sensitivity for several previously unexplored channels and discuss their dependence on center-of-mass energy and the lepton flavor interacting with the sterile neutrino.

Our study builds upon the earlier analyses that we summarize here.   
In Ref.~\cite{Magill:2018jla}, based on the LEP cross section limits \cite{DELPHI:1996drf}, constraints on the transition neutrino magnetic moment were derived using analytical methods; in Ref.~\cite{Zhang:2023nxy}, LEP was revisited and projections for CEPC \cite{CEPCStudyGroup:2018ghi} were derived. In Ref.~\cite{Ovchynnikov:2023wgg}, the displaced vertex signature arising from sterile neutrino decay was explored for several colliders. Ref.~\cite{Barducci:2024kig} (see also \cite{Frigerio:2024pvc} for the SM neutrino magnetic moment) focused on the MuC \cite{Accettura:2023ked} and introduced a channel featuring fusion of $\gamma/Z$ and a neutrino. The fusion of $W^\pm$ boson and a charged lepton, as well as vector boson fusion process, are also important at high-energy lepton colliders for the model under consideration, as we will demonstrate in this work.

This work is organized as follows. In \cref{sec:theory}, we introduce the theoretical framework used to discuss relevant sterile neutrino production and decay channels. \Cref{sec:analysis} outlines the details of our analysis, with additional information elaborated in the Appendix. \Cref{sec:results} presents the main results, and \cref{sec:conclusions} provides our conclusions.


\section{The model}
\label{sec:theory}
\noindent
In this work we consider the active-to-sterile neutrino transition magnetic moment which arises from the following dimension-6 dipole operators, defined above the electroweak scale \cite{Magill:2018jla,Brdar:2020quo}
\begin{align}
    \mathcal{L}  \supset  \frac{c_B}{\Lambda^2} g' B_{\mu \nu} \overline{L} \Tilde{H} \sigma^{\mu \nu} N + \frac{c_W}{\Lambda^2} g W_{\mu \nu}^{a}  \overline{L} \sigma^a \Tilde{H} \sigma^{\mu \nu} N + \text{h.c.}\,.
    \label{eq:dim6Lagrangian}
\end{align} 
Here, $H$, $L$, and $N$ are the Higgs field, the lepton doublet, and the sterile neutrino, respectively, and $N$ is assumed to be of Majorana nature. Further, $g$ and $g'$ are $SU(2)_\text{L}$ and $U(1)_Y$ gauge couplings, $\Lambda$ is the energy scale at which UV complete model is realized, $c_B$ and $c_W$ are  Wilson coefficients, $\sigma^a$ represents Pauli matrices, $\Tilde{H}=i\sigma^2 H^*$ is the conjugate Higgs field and $W^a_{\mu \nu}$, $B_{\mu \nu}$ are field strength tensors. It is crucial to note that in weakly coupled UV theories, these dimension-6 operators are typically generated at one-loop level \cite{Craig:2019wmo}, which suppresses the Wilson coefficients $c_B$ and $c_W$ by a factor of $\mathcal{O}(1/16\pi^2)$. However, Wilson coefficients of $\mathcal{O}(1)$ can also be generated, such as in strongly coupled UV theories.

Below the electroweak symmetry breaking scale the above terms can be written as
\begin{align}
    \mathcal{L}  \supset & v\frac{c_W}{\Lambda^2} g W^-_{\mu \nu} \bar{\ell}_L \sigma^{\mu \nu} N + \frac{v}{\sqrt{2}} \left(\frac{c_B}{\Lambda^2} g' \cos{\theta_w} + \frac{c_W}{\Lambda^2} g \sin{\theta_w} \right) F_{\mu \nu}  \bar{\nu}_L \sigma^{\mu \nu} N \nonumber\\ 
    & +  \frac{v}{\sqrt{2}} \left(-\frac{c_B}{\Lambda^2} g' \sin{\theta_w} +  \frac{c_W}{\Lambda^2} g \cos{\theta_w} \right) Z_{\mu \nu} \bar{\nu}_L \sigma^{\mu\nu} N \nonumber\\ 
    &  -i v\frac{c_W}{\Lambda^2} g^2 \sin \theta_w A_{\nu} W^- _\mu \bar{\ell}_L \sigma^{\mu \nu} N 
    -i v\frac{c_W}{\Lambda^2} g^2 \cos \theta_w Z_{\nu} W^- _\mu \bar{\nu}_L \sigma^{\mu \nu} N \nonumber\\ 
    & + i v\frac{c_W}{\sqrt{2}\Lambda^2} g^2 W^+_\nu W^-_\mu \bar{\nu}_L \sigma^{\mu \nu} N + \text{h.c.}\,, 
    \label{eq:da}
\end{align}
where now charged lepton ($\ell$) and neutrino ($\nu$) fields appear as well as the field strength tensors associated to the photon, $Z$ and $W$ boson, with $W^\pm_{\mu\nu}\equiv \partial_{\mu}W^\pm_{\nu}-\partial_{\nu}W^\pm_{\mu} $ and $Z_{\mu\nu}\equiv \partial_{\mu}Z_{\nu}-\partial_{\nu}Z_{\mu} $. Here, $v=246$ GeV is the vacuum expectation value of the Higgs field and $\theta_w$ is the weak mixing angle.   

Using the following abbreviations
\begin{align}
    d_{\gamma}  = & \frac{ev}{\sqrt{2} \Lambda^2} \left( c_B + c_W \right)\,,\nonumber\\
    d_Z  = & \frac{ev}{\sqrt{2} \Lambda^2} \left( c_W \cot{\theta_w} - c_B \tan{\theta_w}\right)\,,\nonumber\\
    d_W  = & \frac{ev}{ \Lambda^2 \sin{\theta_w}} c_W \,,
      \label{eq:dip}
\end{align}
where $e=g\sin{\theta_w}=g'\cos{\theta_w}$ was used, 
we can express the Lagrangian as 
 \begin{align}
 \mathcal{L} \supset &\, d_\gamma F_{\mu \nu}  \bar{\nu}_L \sigma^{\mu \nu} N + d_W W^-_{\mu \nu} \bar{\ell}_L \sigma^{\mu \nu} N  +  d_Z  Z_{\mu \nu} \bar{\nu}_L \sigma^{\mu\nu} N \nonumber\\ 
 &  -i d_W g W^-_\mu \left( \sin \theta_w A_{\nu}  +  \cos \theta_w Z_{\nu} \right)  \bar{\nu}_L \sigma^{\mu \nu} N + i \frac{d_W}{\sqrt{2}} g W^-_\mu W^+_\nu  \bar{\nu}_L \sigma^{\mu \nu} N  + \text{h.c.}\,. 
 \label{eq:lag}
\end{align}
The first term in \cref{eq:lag} describes the active-to-sterile neutrino transition magnetic moment. For brevity, we have omitted flavor indices in the above equations. We will assume that sterile neutrino interacts in a flavor-specific way with $d_\gamma^\alpha$ denoting transition magnetic moment between sterile neutrino and a neutrino of flavor $\alpha$ ($\alpha=e$, $\mu$, $\tau$). When presenting results, we will emphasize which generation of leptons the derived sensitivities apply to. 

\subsection{Sterile neutrino production}
\label{subsec:production}

\begin{figure}[t!]
    \centering
        \includegraphics[scale=0.85]{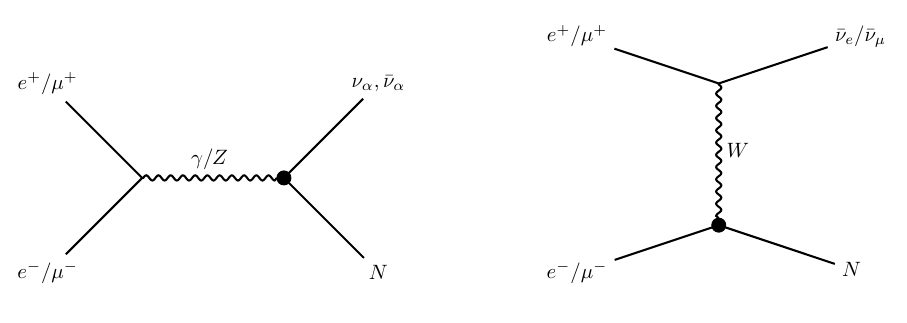}
    \caption{Feynman diagrams corresponding to the s-channel (left panel) and t-channel (right panel) 2-2 process for sterile neutrino production at lepton colliders. In addition to sterile neutrino, $N$, SM neutrino is produced as well. The black blobs represent the insertion of the dimension-6 operators (see \cref{eq:da}). }
    \label{fig:diag1}
\end{figure}

\begin{figure}[t!]
    \centering
        \includegraphics[width=\textwidth]{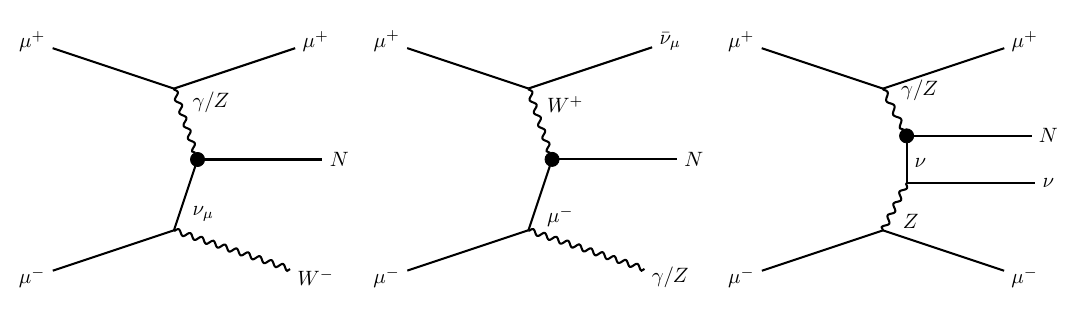}
    \caption{Representative diagrams for 2-3 (left and middle panels) and 2-4 (right panel) processes for sterile neutrino production at the proposed MuC.}
    \label{fig:diag2}
\end{figure}

\noindent
The Feynman diagrams corresponding to the 2-2 process for sterile neutrino production are shown in \cref{fig:diag1}. In the s-channel, together with sterile neutrino, neutrino and antineutrino of any flavor $\alpha$ can be produced. The analytical results for the cross section, calculated by taking into account both s-channel and t-channel diagrams, are shown in \cref{app:xsec}. 

At high-energy colliders, in addition to the 2-2 process, other channels become relevant as well. Specifically, $\gamma/Z$-$\nu$  \cite{Barducci:2024kig} and $W$-$\ell$ fusion lead to the 2-3 processes, $\ell^+ \ell^- \rightarrow N W \ell $ and $\ell^+ \ell^- \rightarrow N  (Z/\gamma) \nu$, shown in the left and middle panels of \cref{fig:diag2} for $\mu^+ \mu^-$ initial states. As can be seen from these two diagrams, in such realizations, sterile neutrinos can only be produced if they interact with the charged lepton and neutrino flavor that matches the flavor of the initial states. Moreover, there are vector boson fusion channels (2-4 processes), and a representative diagram for these processes is shown in the right panel of \cref{fig:diag2}. It should be noted that this diagram contributes regardless of whether the lepton flavor interacting with the sterile neutrino matches the flavor corresponding to the initial states. 

\begin{figure}[h!]
  \centering
  \begin{tabular}{cc}
    $\quad \quad $ (a) & $\quad \quad$  (b) \\
    \includegraphics[width=0.45\textwidth]{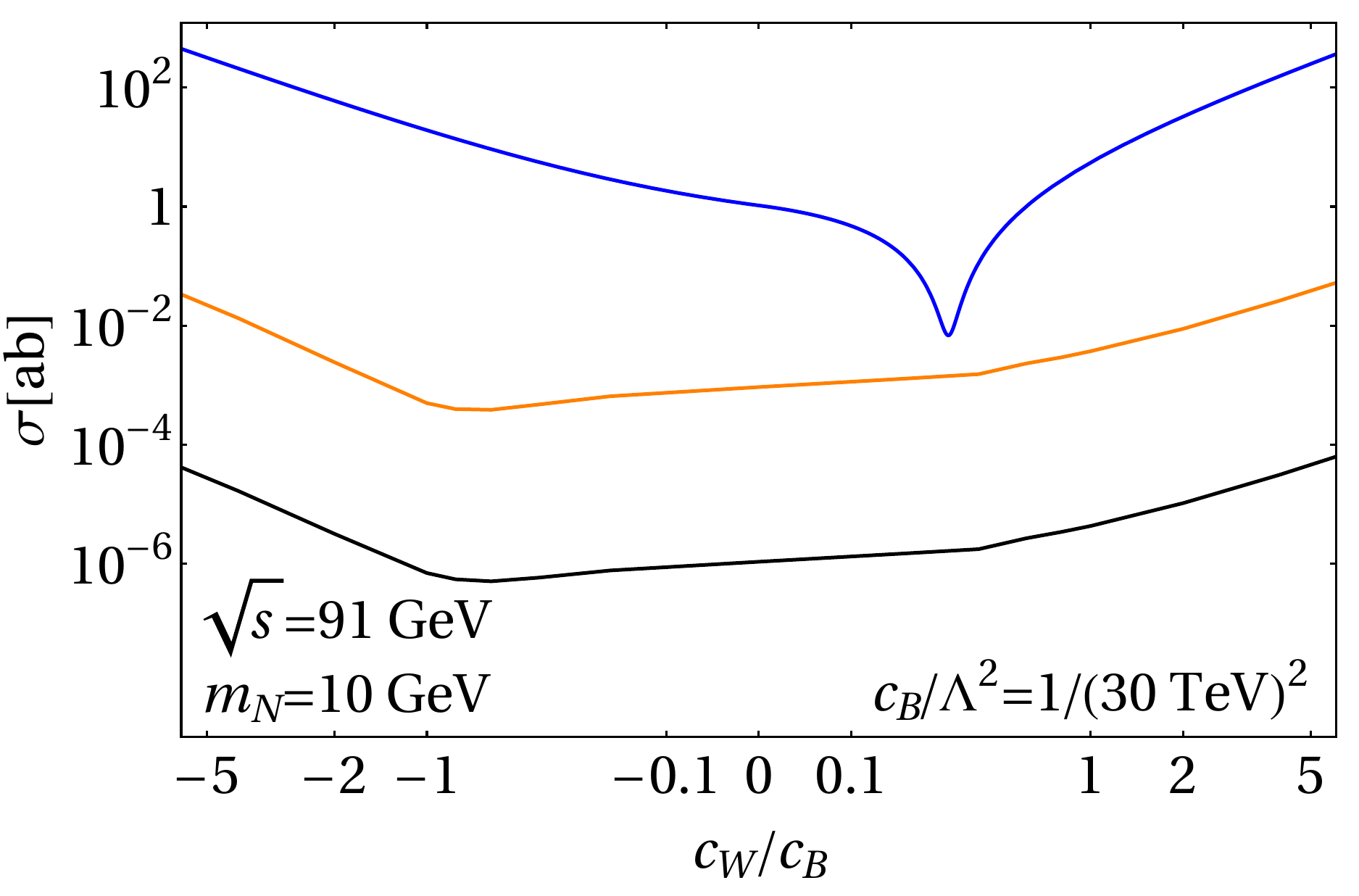} 
    \label{fig:3TeV} &
    \includegraphics[width=0.45\textwidth]{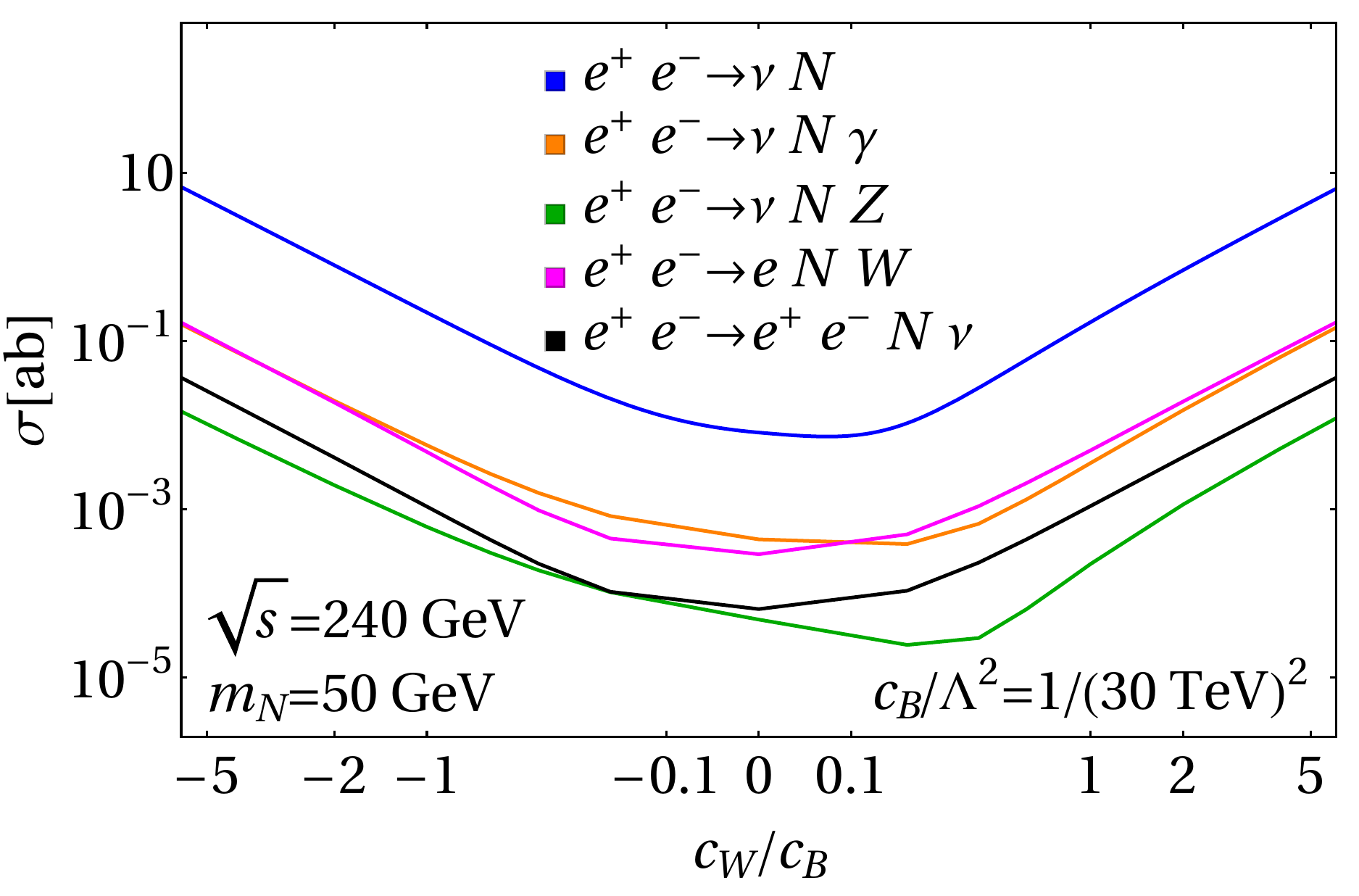} 
    \label{fig:10TeV}\\ 
  \end{tabular}

  \begin{tabular}{cc}
       $\quad  \quad$ (c) & $\quad \quad$  (d) \\
    \includegraphics[width=0.45\textwidth]{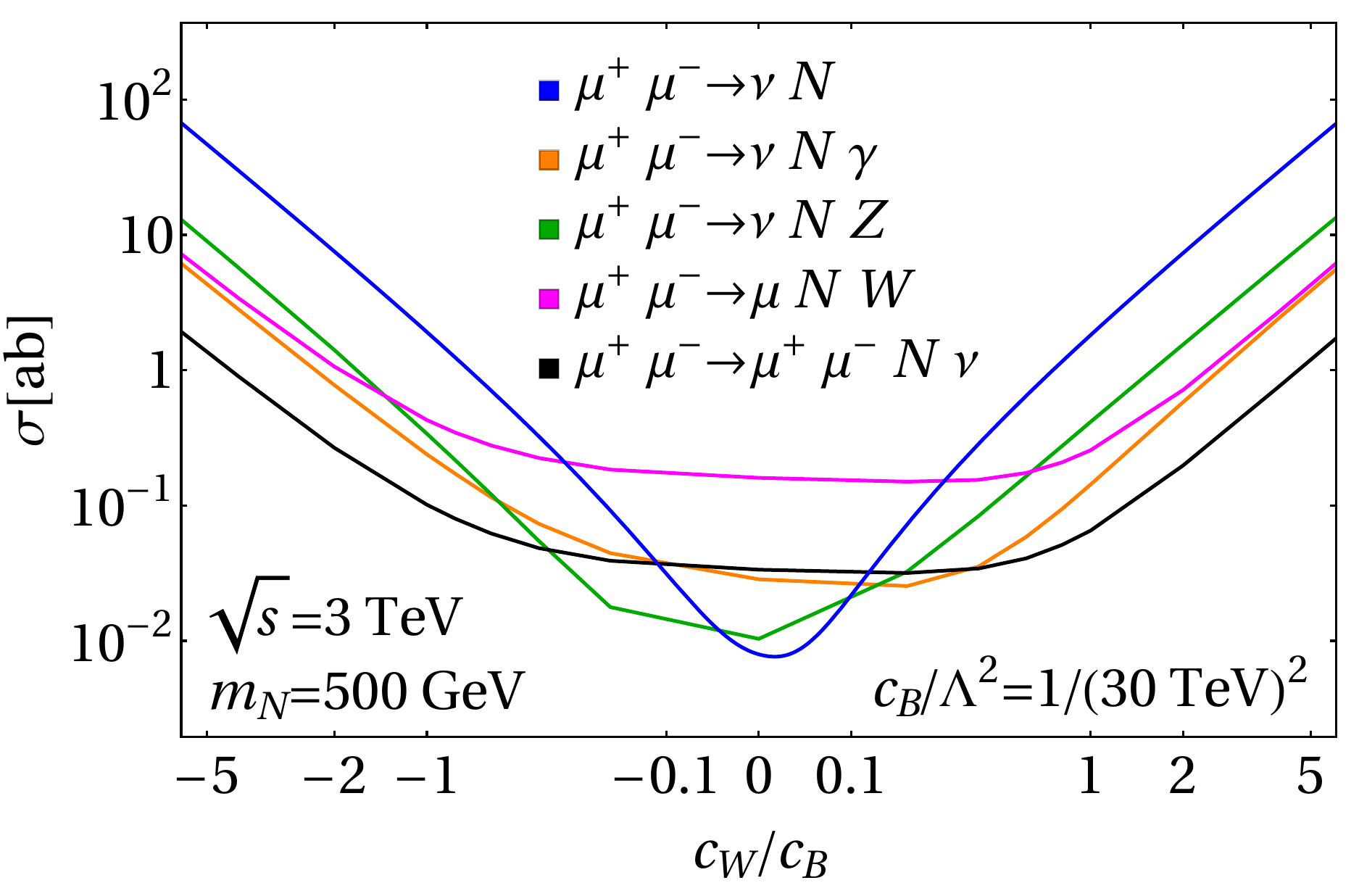}
    \label{fig:240GeV}  &
    \includegraphics[width=0.45\textwidth]{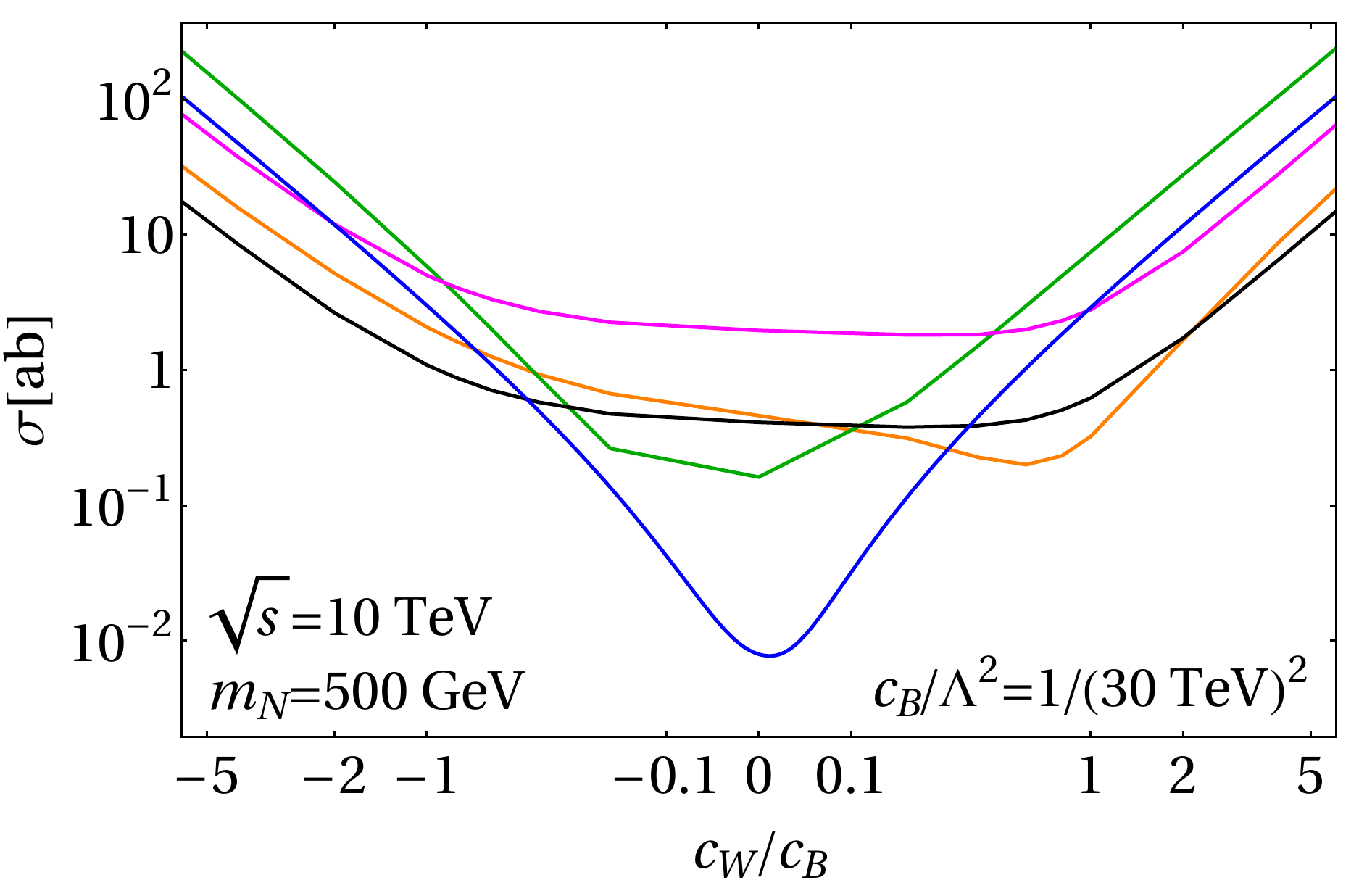} 
    \label{fig:91GeV}\\
  \end{tabular}
    \caption{Upper (lower) panels show cross sections  with $c_B/\Lambda^2 = 1/(30 \text{ TeV})^{2}$ (after generator-level cuts discussed in \cref{sec:Event cuts}) for various sterile neutrino production channels, evaluated for $e^+e^-$ ($\mu^+\mu^-$) collider at low (high) center-of-mass energy $\sqrt{s}$. Notice the relative suppression of the cross sections corresponding to the 2-3 and 2-4 processes (colors other than blue) at smaller $\sqrt{s}$ as well as their relevance for $\sqrt{s}\gtrsim 1$ TeV. The narrow  dip in panel (a) arises from absence of sterile neutrino production via the $Z$ resonance for $d_Z=0$  $(c_W/c_B=0.286)$. }
\label{fig:xsecBRplot2} 
\end{figure}

In \cref{fig:xsecBRplot2}, we present the cross section for various sterile neutrino production channels, including aforementioned 2-2, 2-3 and 2-4 processes. \newtext{We calculated these cross sections from \texttt{MadGraph5\_aMC@NLO} \cite{Alwall:2014hca} using the UFO model file generated after implementing \cref{eq:dim6Lagrangian} in \texttt{FeynRules} \cite{Alloul:2013bka}. } In each panel, the cross section is shown as a function of $c_W/c_B$ and for a specific value of sterile neutrino mass, $m_N$, and center-of-mass energy $\sqrt{s}$. It is evident from the lower panels that 2-3 and 2-4 processes become relevant for $\mathcal{O}(\text{TeV})$ values of center-of-mass energy. Note that panel (a) in 
\cref{fig:xsecBRplot2} contains only 3 lines since, at $\sqrt{s}=91$ GeV and with $m_N=10$ GeV, production of $W$ and $Z$ boson is kinematically forbidden. 
The narrow  dip in this  panel arises from absence of sterile neutrino production via the $Z$ resonance.

\par

\subsection{Sterile neutrino decay}
\label{subsec:decay}
\begin{figure}[t!]
  \centering
  \begin{tabular}{cc}
    $\quad \quad $ (a) & $\quad \quad$  (b) \\
    \includegraphics[width=0.48\textwidth]{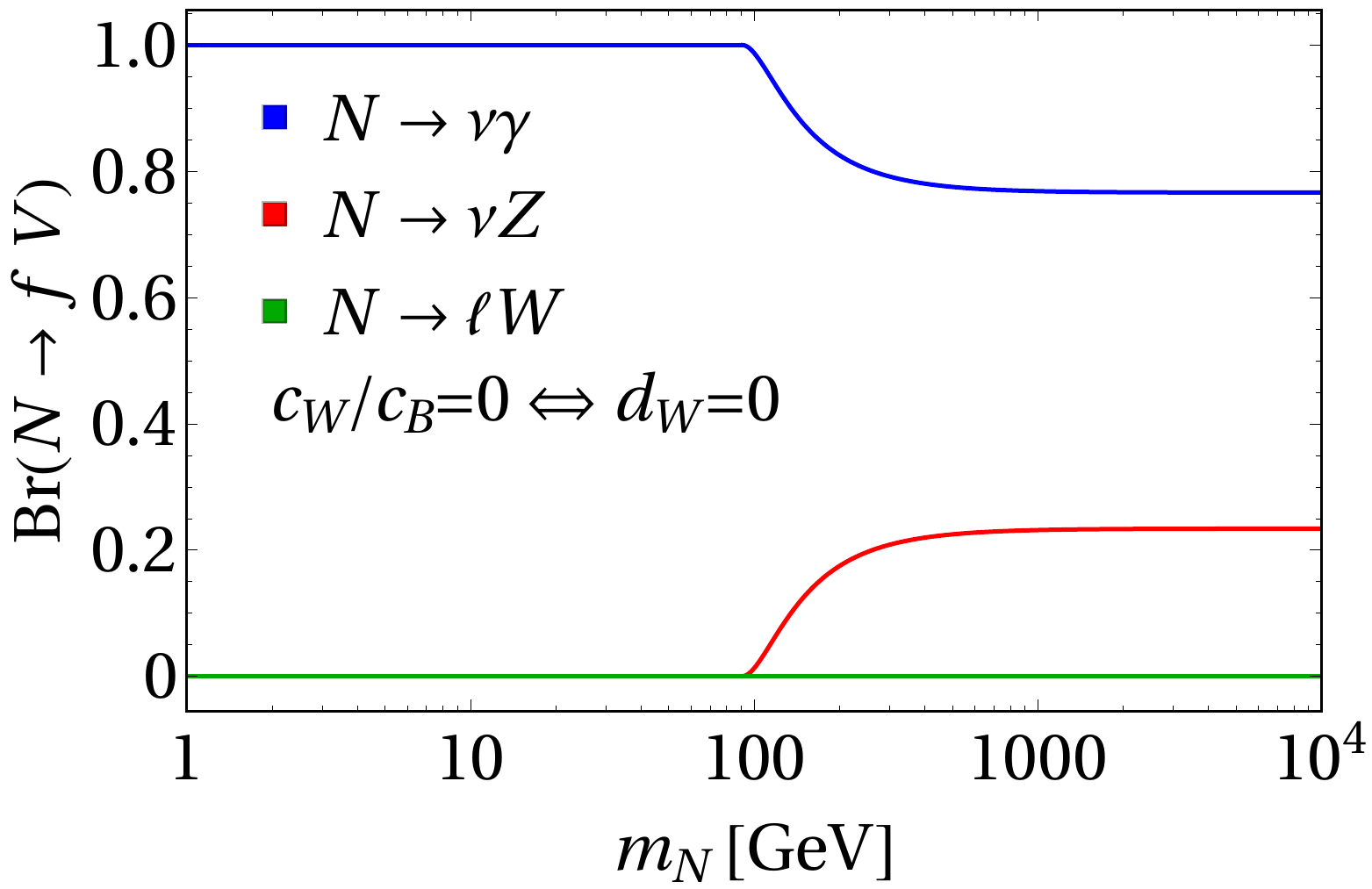}  &
    \includegraphics[width=0.48\textwidth]{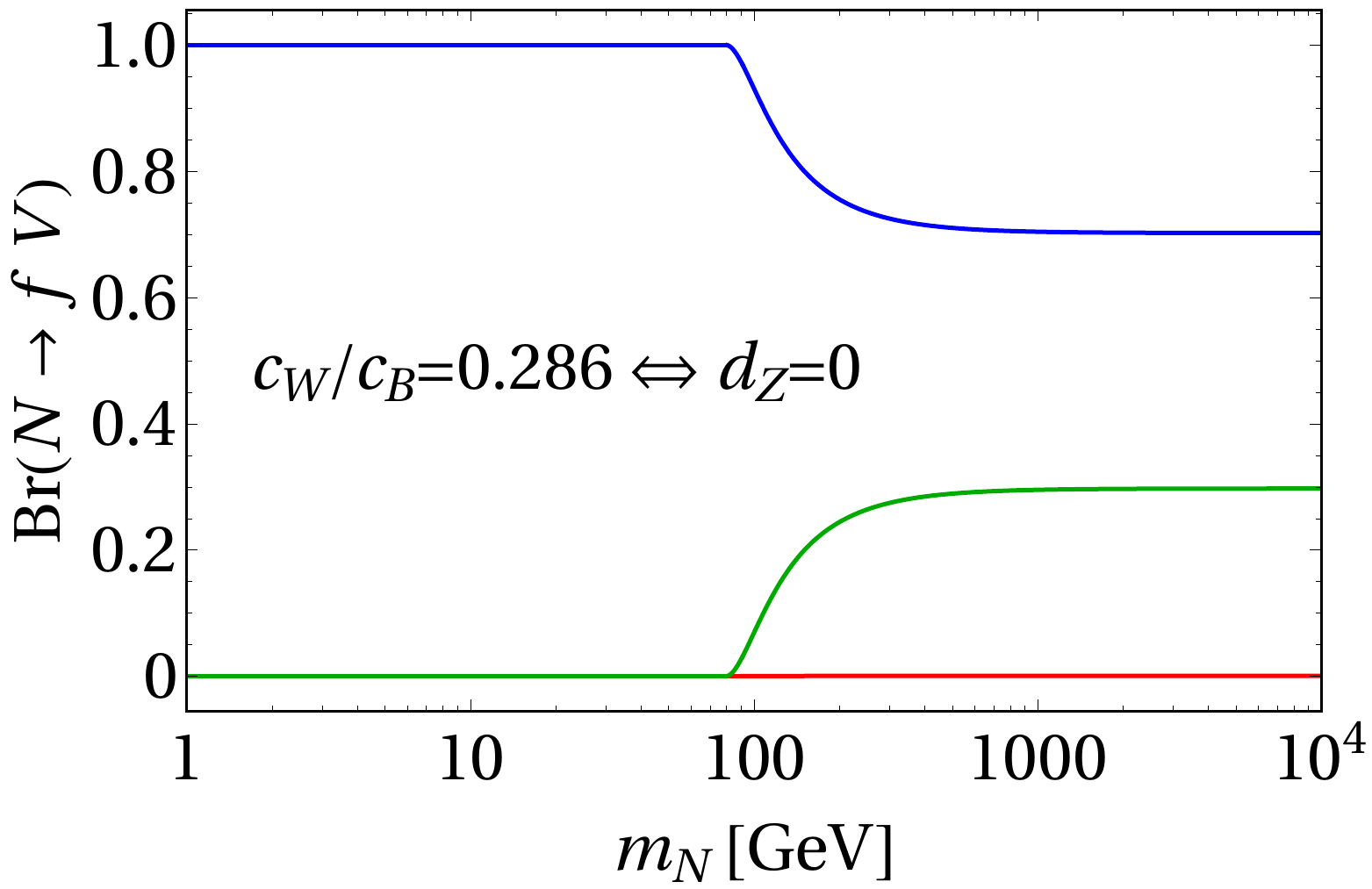}  \\ 
  \end{tabular}

  \begin{tabular}{cc}
       $\quad  \quad$ (c) & $\quad \quad$  (d)\\
    \includegraphics[width=0.48\textwidth]{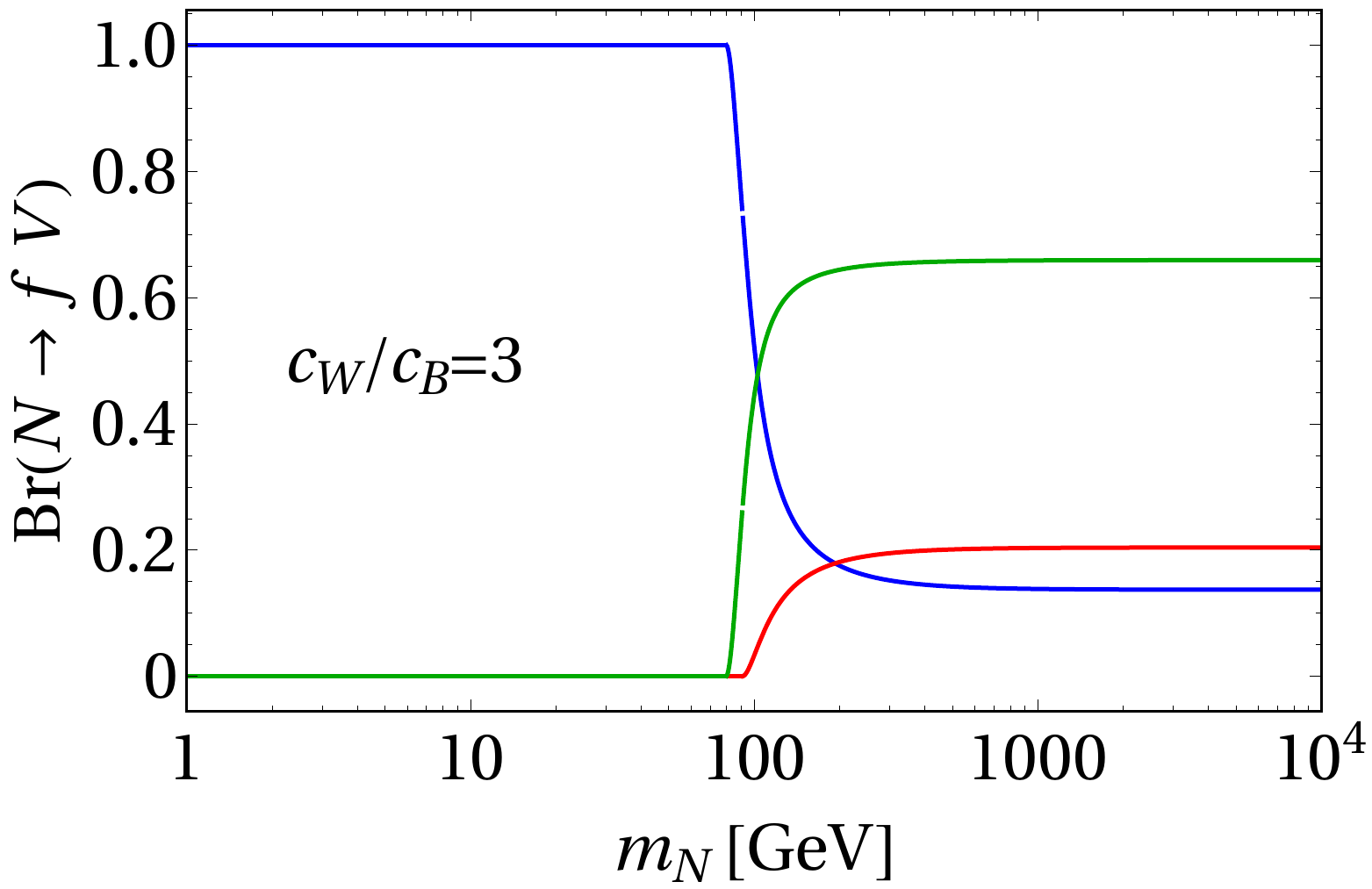} &
    \includegraphics[width=0.48\textwidth]{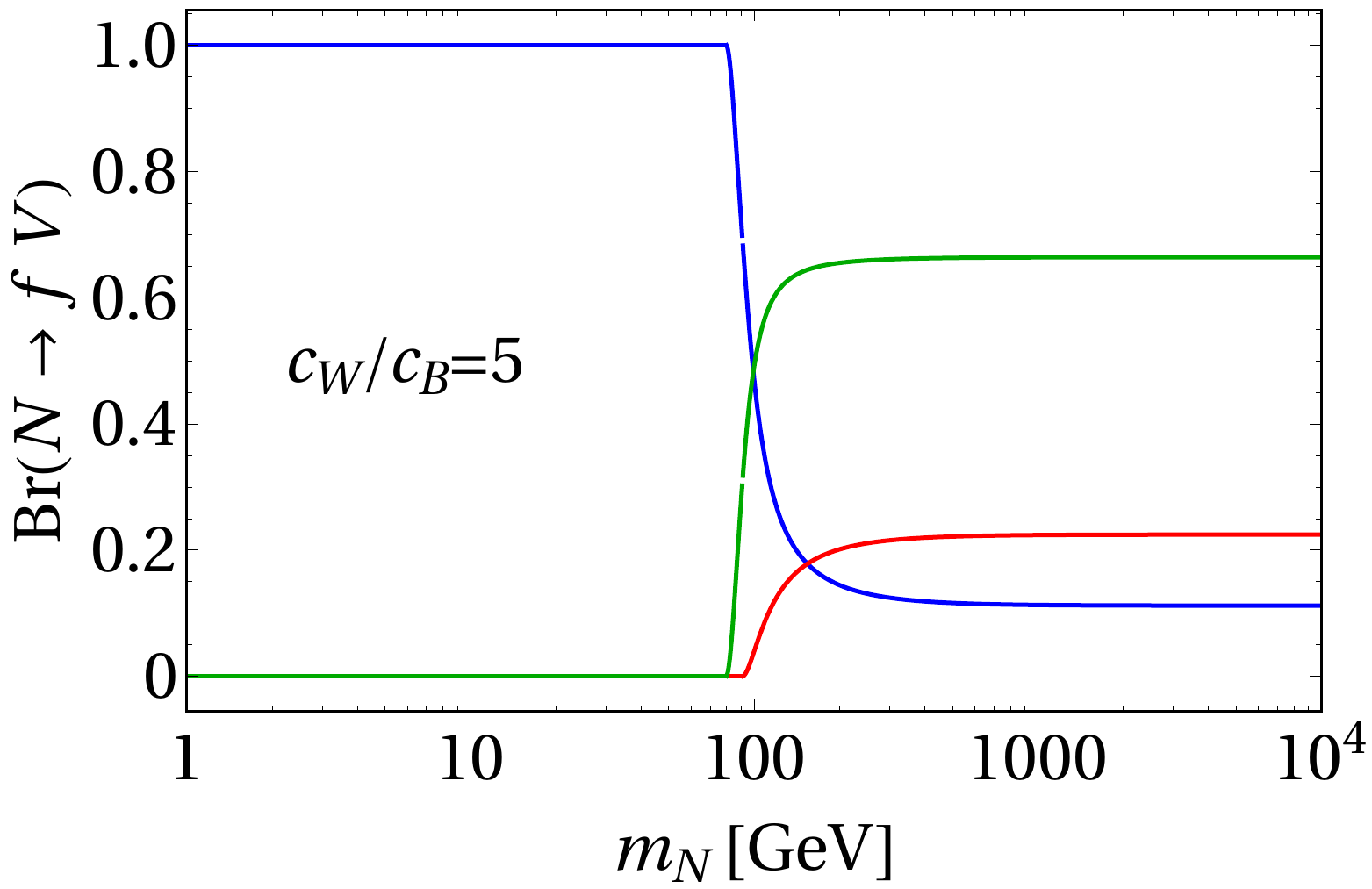}\\
  \end{tabular}
    \caption{Branching ratios for $N\rightarrow\nu\gamma$, $N\rightarrow\nu Z$ and $N\rightarrow \ell^{\pm}W^{\mp}$ decay channels are shown for four benchmark points considered in this work.}
    \label{fig:xsecBRplot3}
\end{figure}

\begin{figure}[t!]
  \centering
  \begin{tabular}{cc}
    $\quad \quad $ (a) & $\quad \quad$  (b) \\
    \includegraphics[width=0.48\textwidth]{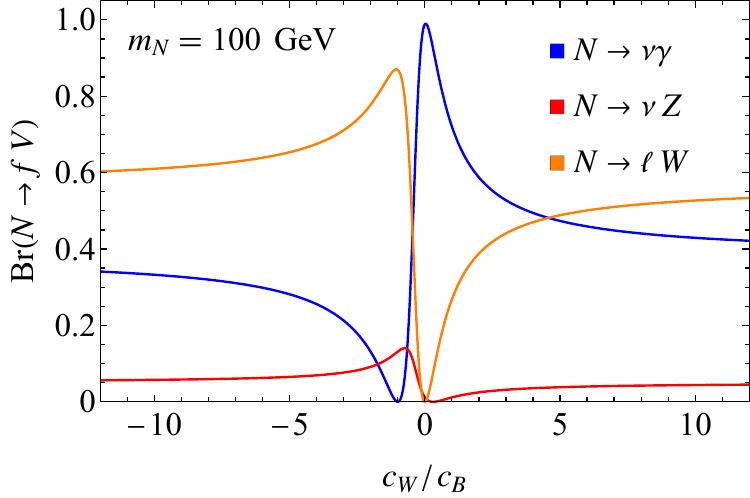}  &
    \includegraphics[width=0.48\textwidth]{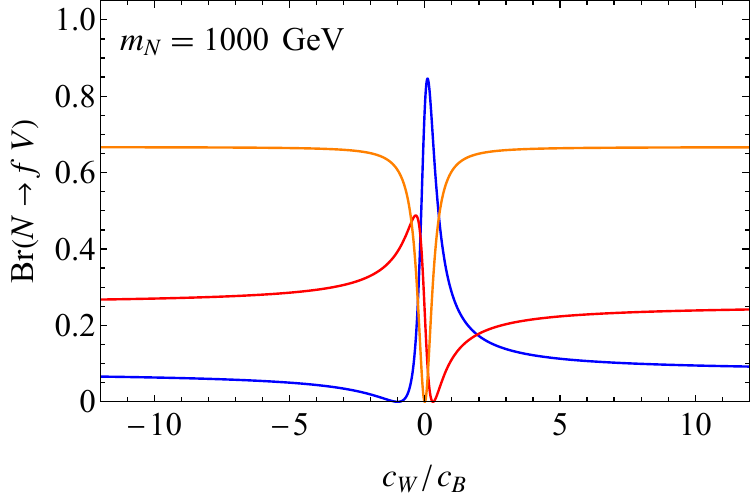}  \\ 
  \end{tabular}

    \caption{\newtext{Branching ratios for $N\rightarrow\nu\gamma$, $N\rightarrow\nu Z$ and $N\rightarrow \ell^{\pm}W^{\mp}$ decay channels are shown as a function of $c_W/c_B$ for two different sterile neutrino masses $m_N=100\text{ GeV}$ and $m_N=1000\text{ GeV}$.}}
    \label{fig:xsecBRplot_BRmN}
\end{figure}

\noindent
There are three sterile neutrino decay channels, $N\rightarrow\nu\gamma$, $N\rightarrow\nu Z$ and $N\rightarrow \ell^{\pm}W^{\mp}$, with the following decay rates
\begin{align}
    \Gamma_{N\rightarrow\nu\gamma} &= \frac{d_\gamma^2 m_N^3}{2\pi}\,, \nonumber \\
    \Gamma_{N\rightarrow\nu Z} &= \frac{ d_Z^2 }{4\pi m_N^3}\left(m_N^2-m_Z^2\right)^2\left(2m_N^2+m_Z^2\right)\,, \nonumber \\
    \Gamma_{N\rightarrow \ell^{\pm}W^{\mp}}  &= \frac{ d_W^2 }{4 \pi m_N^3}\sqrt{ \left( \left(m_{\ell}-m_W\right) ^2 -m_N^2 \right) \left(\left(m_{\ell}+m_W\right) ^2 -m_N^2 \right) }\nonumber \\ 
     &  \times \left(2\left(m_{\ell}^2-m_N^2\right)^2-m_W^2\left(m_{\ell}^2+m_N^2\right)-m_W^4\right)\,.
    \label{eq:decay_width}
\end{align}
Here, $m_\ell$, $m_W$ and $m_Z$ are charged lepton, $W$ boson and $Z$ boson mass, respectively. When $m_N \gg 100$ GeV, the branching ratio for a particular decay channel is approximately equal to
\begin{align}
    \text{Br}(N \to f V) \simeq \frac{d^2_V}{d_{\gamma}^2 + d_{W}^2 + d_Z^2}\,.
\label{eq:BR}
\end{align}
Here, $V = \gamma, Z$, $W$ and $f$ represents charged leptons or neutrinos. The form of \cref{eq:BR} can be understood by recognizing that for large $m_N$ all the decay products can be approximated to be massless and the only difference among the decay modes arises from the corresponding couplings, which are independent of $m_N$. 
The branching ratios are shown in \cref{fig:xsecBRplot3} for four representative benchmark points. We note that $\text{Br}\,(N\rightarrow \ell^{\pm} W^{\mp}) = 0$ for any value of $m_N$ in panel (a) as a result of the $d_W = 0$ benchmark point, and analogously, $\text{Br}\,(N\rightarrow\nu Z) = 0$ in panel (b) due to the $d_Z = 0$ benchmark point. \newtext{In \cref{fig:xsecBRplot_BRmN}, we also present the branching ratios as a function of the ratio of the Wilson coefficients, $c_W/c_B$, for two values of $m_N$.}

\par

\par

\section{Analysis}
\label{sec:analysis}
\noindent
CEPC \cite{CEPCStudyGroup:2018ghi}, FCC-ee \cite{FCC:2018evy}, CLIC \cite{Roloff:2018dqu} and MuC \cite{Accettura:2023ked} are proposed lepton collider experiments in the focus of this work. The former three are $e^+ e^-$ colliders while MuC will feature muon collisions. We list the center-of-mass energy and integrated luminosity, $\mathcal{L}$, for all these experiments in \cref{tab:exp_CMenergy_lumi} where we also include former lepton collider LEP \cite{Assmann:2002th}. The details of our analysis for all of these experiments are outlined below; they form the basis for the results presented in \cref{sec:results} where we show the sensitivity reach for the neutrino transition magnetic moment.

\begin{table}[h!]
    \centering
    \begin{tabular}{|c|c|c|}
        \hline
        Experiment & $\sqrt{s}$ (GeV) & $\mathcal{L}$ ($\text{ab}^{-1}$) \\
        \hline
        \hline
        LEP \cite{Assmann:2002th} & 91 & $2\times10^{-4}$ \\
        \hline
        CEPC \cite{CEPCStudyGroup:2018ghi}/FCC-ee \cite{FCC:2018evy}  & 240 & $5.6$\\
        \hline
        CLIC \cite{Roloff:2018dqu} & $3\times10^3$ & $3$ \\
        \hline
        MuC–3 \cite{Accettura:2023ked} & $3\times10^3$ & $3$ \\
        \hline
        MuC–10 \cite{Accettura:2023ked} & $10\times10^3$ & $10$ \\
        \hline
    \end{tabular}
    \caption{The considered collider experiments, their expected center-of-mass energy $\sqrt{s}$ and integrated luminosity $\mathcal{L}$.} 
    \label{tab:exp_CMenergy_lumi}
\end{table}
We produced a \texttt{FeynRules} \cite{Alloul:2013bka} model file featuring the dimension-6 Lagrangian from \cref{eq:dim6Lagrangian} which allowed us to employ \texttt{MadGraph5\_aMC@NLO} \cite{Alwall:2014hca} for event generation. For the signal events, we consider sterile neutrino production processes discussed in \cref{subsec:production} and incorporate branching ratios for $N$ decay. Specifically, by combining five processes shown in \cref{fig:xsecBRplot2} with the $N \rightarrow \nu \gamma$ and $N \rightarrow \ell^{\pm} W^{\mp}$ decay processes, we arrive at 10 channels that are listed in \cref{tab:all_channels}. In this table we also abbreviate all considered channels; note that the brackets in the second column indicate the decay channels. We choose not to study $N \rightarrow \nu Z$, as this is not the leading decay process for any of the benchmark points shown in \cref{fig:xsecBRplot3}. The initial states quoted in the table are $\mu^+\,\mu^-$, however, note that the channels are analogous for $e^+e^-$ colliders. 

\begin{table}[h!]
    \centering
    \begin{tabular}{|c|c|c|}
    \hline
    & Label & Signal channel \\
    \hline
    \hline
    2-2 &0-$\gamma$ & $\mu^+ \mu^- \rightarrow \nu N (\nu \gamma)$ \\
    \hline
    \multirow{3}{*}{2-3}&$\gamma$-$\gamma$ & $\mu^+ \mu^- \rightarrow \nu \gamma N (\nu \gamma)$ \\
    \cline{2-3}
    &$Z$-$\gamma$ & $\mu^+ \mu^- \rightarrow \nu Z N (\nu \gamma)$ \\
    \cline{2-3}
   & $W$-$\gamma$ & $\mu^+ \mu^- \rightarrow \mu^{\pm} W^{\mp} N (\nu \gamma)$ \\ \hline
   2-4 & $2\ell$-$\gamma$& $\mu^+ \mu^- \rightarrow \nu \ell^{\pm}_1 \ell^{\mp}_2 N (\nu \gamma)$
    \\
    \hline
    \hline
     2-2 & 0-$W$ & $\mu^+ \mu^- \rightarrow \nu N (\mu^\pm W^\mp)$ \\
    \hline
    \multirow{3}{*}{2-3}&$\gamma$-$W$ & $\mu^+ \mu^- \rightarrow \nu \gamma N (\mu^{\pm} W^{\mp})$ \\
    \cline{2-3}
      &$Z$-$W$ & $\mu^+ \mu^- \rightarrow \nu Z N (\mu^{\pm} W^{\mp})$ \\
    \cline{2-3}
    &$W$-$W$ & $\mu^+ \mu^- \rightarrow \mu^{\pm} W^{\mp} N (\mu^{\pm} W^{\mp})$ \\ \hline
    2-4&$2\ell$-$W$& $\mu^+ \mu^- \rightarrow \nu \ell^{\pm}_1 \ell^{\mp}_2 N (\mu^{\pm} W^{\mp})$
    \\
    \hline
    \end{tabular}
    \caption{Signal channels employed in our analysis. The products from sterile neutrino decay are shown in brackets and, in order to simplify the notation, we use $\nu$ for both neutrinos and antineutrinos. Only the first two channels will be studied for CEPC, FCC-ee, and LEP  $e^+ e^-$ colliders while all 10 channels will be scrutinized for MuC and CLIC.}
    \label{tab:all_channels}
\end{table}

\newtext{We focus on scenarios in which sterile neutrinos decay promptly within the detector volume. The characteristic distance $L_D$ traveled by a sterile neutrino during its lifetime is given by
\begin{equation}
    L_D = \frac{1}{\Gamma_{\text{tot}}} \frac{E_N}{m_N} \sqrt{1 - \left( \frac{m_N}{E_N} \right)^2},
\end{equation}
where $ m_N $, $ E_N $, and $ \Gamma_{\text{tot}} $ denote the sterile neutrino mass, energy, and total decay width, respectively. For representative parameters $ m_N > 10~\text{GeV} $ and a transition magnetic moment $ d_\gamma = 10^{-5}~\text{GeV}^{-1} $, we find $ L_D < 10^{-5}\, \text{m} $. Taking the upper limit of \(4 \times 10^{-5}\,\text{m}\) for the decay length in the prompt decay region at the LHC~\cite{Cepeda_2022} as a reference, the sterile neutrino decays well inside the detector. This is particularly true for sterile neutrinos with larger magnetic moments or heavier masses, where prompt decays would occur even more readily. Our study of prompt decay is complementary to studies that focused on displaced vertex signatures (\emph{e.g.} Ref. \cite{Ovchynnikov:2023wgg}) as well as realizations where sterile neutrinos only decay outside the central detector.  }

The number of signal events in the detector depends on the Wilson coefficients $(c_B, c_W)$ and sterile neutrino mass $m_N$. For all considered benchmark points we have $c_W=\text{const.}\times c_B$ making our scan, for a fixed $m_N$, one-dimensional. 
In this work we perform standard cut-and-count analysis in order to calculate the expected sensitivities. We point out that the behavior of the signal depends on  $m_N$ and therefore we found it favorable to apply $m_N$-dependent cuts. To this end, we divided the considered sterile neutrino mass range into three regions and applied appropriate (and different) cuts in each region. As an example, for $\sqrt{s}=10$ TeV, three signal regions (I, II, III) are obtained by dividing the $m_N$ mass range into ``low'' ($m_N \leq 3000\text{ GeV}$), ``medium'' ($3000\text{ GeV}<m_N<7000\text{ GeV}$) and ``high'' ($7000\text{ GeV} \leq m_N$) mass domain. The values $m_N=1500,\,5000,$ and $8500$ GeV are  representative values (and also the mean values of $m_N$ in these regions) for which we find optimal cuts that are subsequently applied across the respective region. In principle, this analysis can be made more fine-grained; such strategy is most appropriately conducted using machine learning algorithms that are beyond the scope of this work.

\begin{figure}[h!]
  \centering
  \begin{tabular}{cc}
    $\quad \quad $ (a) & $\quad \quad$  (b) \\
    \includegraphics[width=0.48\textwidth]{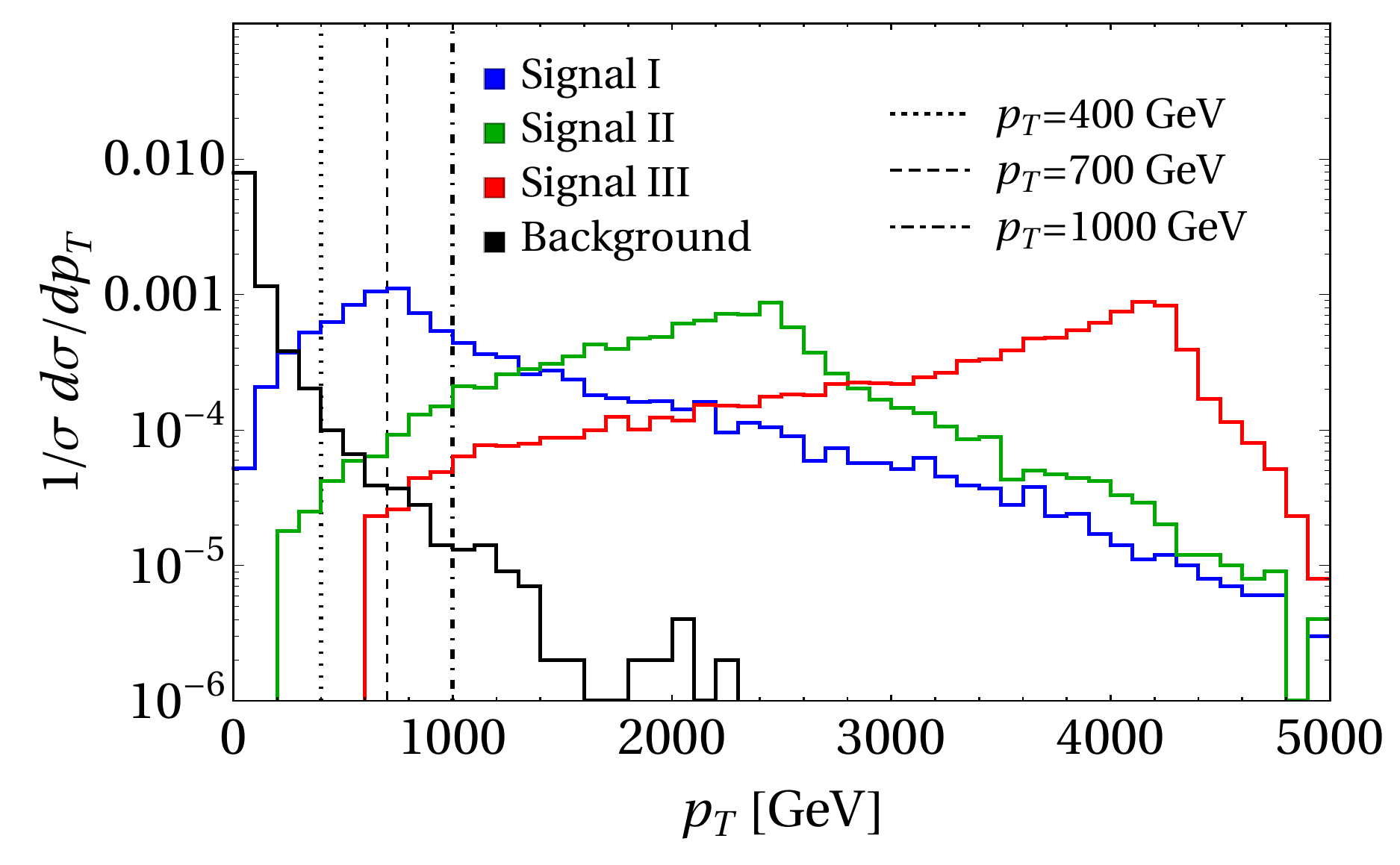}  &
    \includegraphics[width=0.48\textwidth]{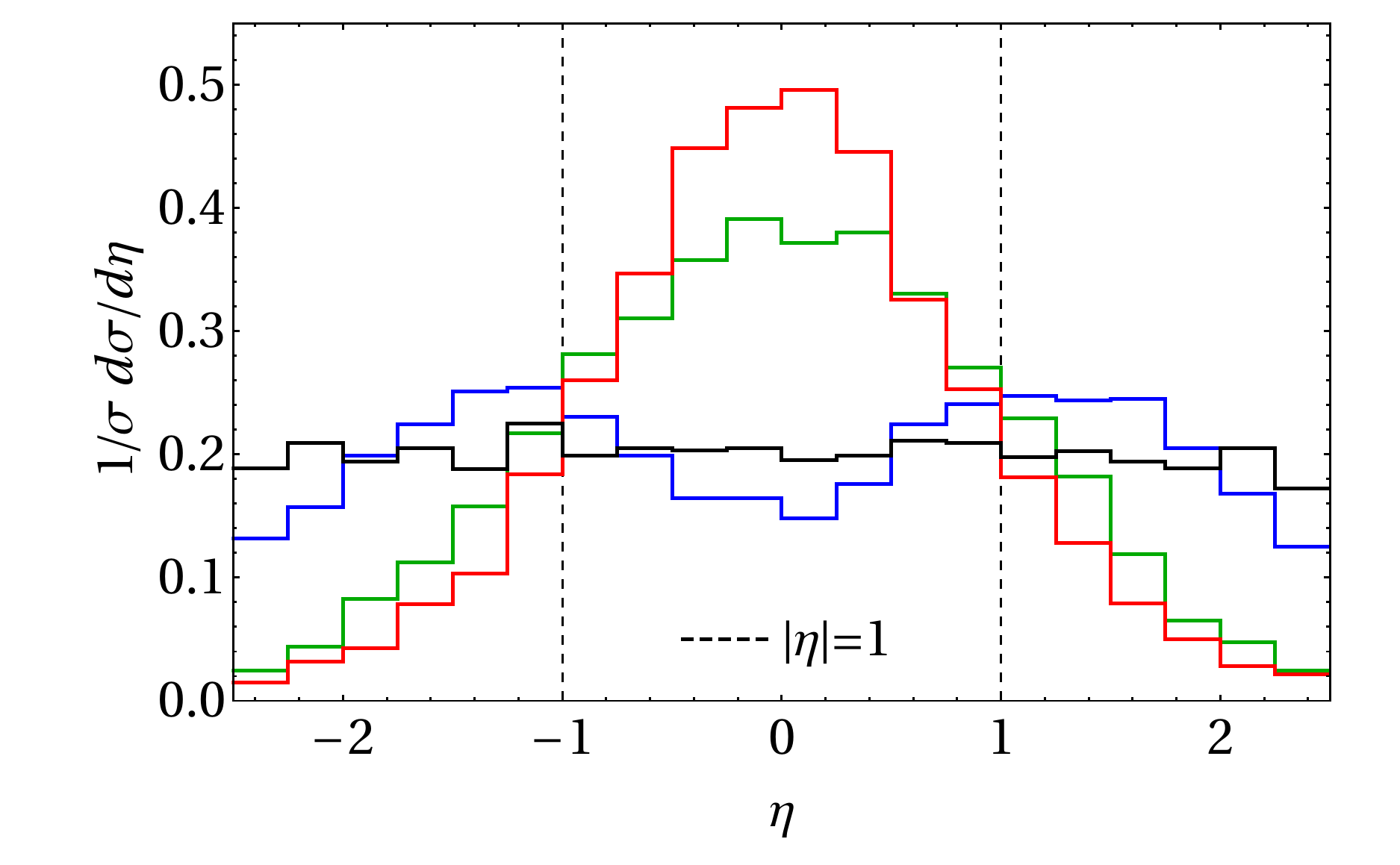}  \\ 
  \end{tabular}
\caption{The figure shows $p_T$ (left panel) and $\eta$ (right panel) distributions of signal and background for 0-$\gamma$ channel at MuC ($\sqrt{s}=10$ TeV). The signal events are generated at $c_W/c_B = 3$ benchmark. The vertical lines indicate the cuts.}
    \label{fig:cut_signal_BG_muonC10TeV1gamma}
\end{figure}

The details on the performed cuts are given in \cref{sec:Event cuts,sec:Cut tables} where we discuss and tabulate all the implemented generator-level and analysis cuts. As an example, in \cref{fig:cut_signal_BG_muonC10TeV1gamma}, we show normalized photon $p_T$ and $\eta$ distributions of signal and background for one particular channel (0-$\gamma$) at MuC. The left panel has three gray, vertical lines at 400, 700, and 1000 GeV that indicate the cuts on the $p_T$ distribution in each of the three $m_N$ regions. These values are obtained by finding $p_T$ for which normalized signal distribution starts dominating over the background distribution. 
For example, when $p_T$ exceeds these values, Signal I (blue), II (green), and III (red) are larger than the background (black), respectively. For the $\eta$ distribution shown in the right panel, we observe that Signal I does not significantly exceed the background in any region of the phase space, while both Signal II and III clearly surpass the background in the $|\eta|<1$ region (dashed line). We impose $|\eta|<1$ as a cut for Signal II and III without applying any $|\eta|$ cut on Signal I.

\par

After performing the cuts, we find the cut efficiency for the signal and background, defined by $\epsilon = N_{\text{cut}}/N_{0}$, where $N_{\text{cut}}$ is the number of events that survive the cuts, and $N_{0}$ is the number of generated events.

The final number of events for signal (S) and background (B) is given by, 
\begin{equation}
S = \sigma_{S} \, \epsilon_{S} \, \mathcal{L}\,, \hspace{1in} B = \sigma_{B} \, \epsilon_{B} \, \mathcal{L}\,.
\label{eq:sig_bg_count}
\end{equation}
Here, $\sigma$ is the generator-level cross-section and $\mathcal{L}$ is the collider luminosity. Subscripts S and B on the cross section and efficiency denote whether these quantities are associated to the signal or background.

The log-likelihood is defined as
\begin{equation}
    -2\log L =  2 \left( N_{\text{exp}} - N_{\text{obs}} + N_{\text{obs}} \log \frac{N_{\text{obs}}}{N_{\text{exp}}} \right)\,.
    \label{eq:chi2}
\end{equation}
Since we are computing exclusion limits, the expected number of events is $N_{\text{exp}}=S+B$, while for the observed number of events we set $N_{\text{obs}}=B$. \cref{eq:chi2} is employed because it yields valid results even in the limit of very small $N_{\text{obs}}$. The 95\% confidence level sensitivities that we will present in \cref{sec:results} correspond to $-2\log {L} = 3.841$. 

\section{Results}
\label{sec:results}
\noindent
The model parameters are $m_N$ as well as $d_\gamma$, $d_W$ and $d_Z$ couplings which depend on two Wilson coefficients, $c_W$ and $c_B$ ( see \cref{eq:dip}). The typical benchmark scenarios considered in the literature are  $d_Z=0 \,(\Leftrightarrow c_W \simeq 0.286\, c_B)$ and $d_W=0 \, (\Leftrightarrow c_W=0)$ \cite{Magill:2018jla,Zhang:2023nxy, Barducci:2024kig}. In this work, however, we consider additional benchmark points, $c_W/c_B=3$ and $c_W/c_B=5$, which will allow us to demonstrate the importance of several new channels introduced in this analysis. We will present results in $m_N$-$d_\gamma^\alpha$ parameter space, since our primary interest lies in the transition magnetic moment $d_\gamma^\alpha$ for flavor $\alpha$.

Before discussing results for the lepton colliders, it is useful to review present LHC limits and upcoming HL-LHC \cite{ZurbanoFernandez:2020cco} expected sensitivities. For LHC, using $36.1 \,\text{fb}^{-1}$, the constraint derived in \cite{Magill:2018jla} for $d_W=0$ benchmark point is in the ballpark of $d_\gamma = 10^{-4}$ GeV$^{-1}$. This allows us to estimate the expected sensitivity for HL-LHC as $d_\gamma \simeq 3 \times 10^{-5}$ GeV$^{-1}$. In this section we will show that smaller $d_\gamma$ can be probed at all considered lepton colliders. 

In what follows, we mainly present results for the scenario where sterile neutrino interacts with a lepton flavor that matches the flavor of colliding leptons. When applicable, we also highlight channels which can be used to derive sensitivity for sterile neutrino interaction with other flavors.

\subsection{LEP}
\label{subsec:LEP}
\noindent
We first reexamine LEP constraints on $d_\gamma^e$, which were previously derived in \cite{Magill:2018jla} based on LEP single photon search \cite{DELPHI:1996drf,OPAL:1994kgw}. Using the procedure described in \cref{sec:analysis}, we performed an independent analysis for active-to-sterile neutrino transition magnetic moment scenario. We considered a single photon in the final state (0-$\gamma$) as well as $\gamma$-$\gamma$ channel (both listed in \cref{tab:all_channels}). Even though the sterile neutrino production cross section for the latter channel is generally not competitive with the 0-$\gamma$ one (see panel (a) in \cref{fig:xsecBRplot2}), our aim was to investigate whether the presence of another photon could aid in reducing the background. Our analyses for both channels are based on the LEP run at the $Z$-pole ($\sqrt{s}=91 \,\text{GeV}$).
 
\begin{figure}[h!]
  \centering
    \includegraphics[width=0.55\textwidth]{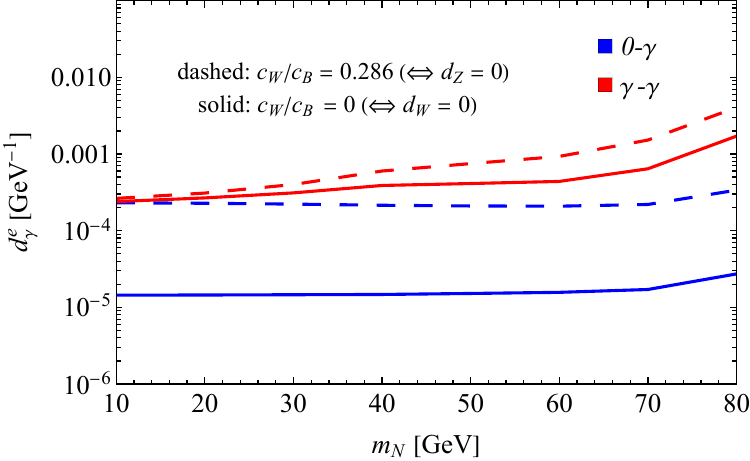} 
  \caption{Derived constraints for the $d_Z=0$ and $d_W=0$ benchmark points at LEP with $\sqrt{s}=91$ GeV.}
  \label{fig:LEP91GeV}
\end{figure}

The signal and background events are subjected to the generator-level and analysis cuts presented in \cref{sec:Event cuts,tab:LEP91}. For the single photon channel 0-$\gamma$, we find comparable limits on the single photon production cross section to those of the LEP collaboration \cite{DELPHI:1996drf,OPAL:1994kgw}. In \cref{fig:LEP91GeV} we show LEP constraints for two benchmark points, $d_Z=0$ and $d_W=0$. For $d_Z=0$ $(d_W=0)$ benchmark we find the limit at around $d_\gamma^e\simeq 10^{-4}$ GeV$^{-1}$ ($d_\gamma^e\simeq 10^{-5}$ GeV$^{-1}$), both arising from the 0-$\gamma$ channel analysis; the dominance of this channel is mainly driven by the larger cross section. Note, however, that for $d_Z=0$ benchmark point around $m_N \simeq \mathcal{O}(10)$ GeV we find the two channels to yield comparable limits; in this parameter space, the difference in the cross section is compensated by signal and background efficiencies. 

\begin{figure}[h!]
  \centering
  \begin{tabular}{cc}
    $\quad \quad $ (a) & $\quad \quad$  (b) \\
    \includegraphics[width=0.48\textwidth]{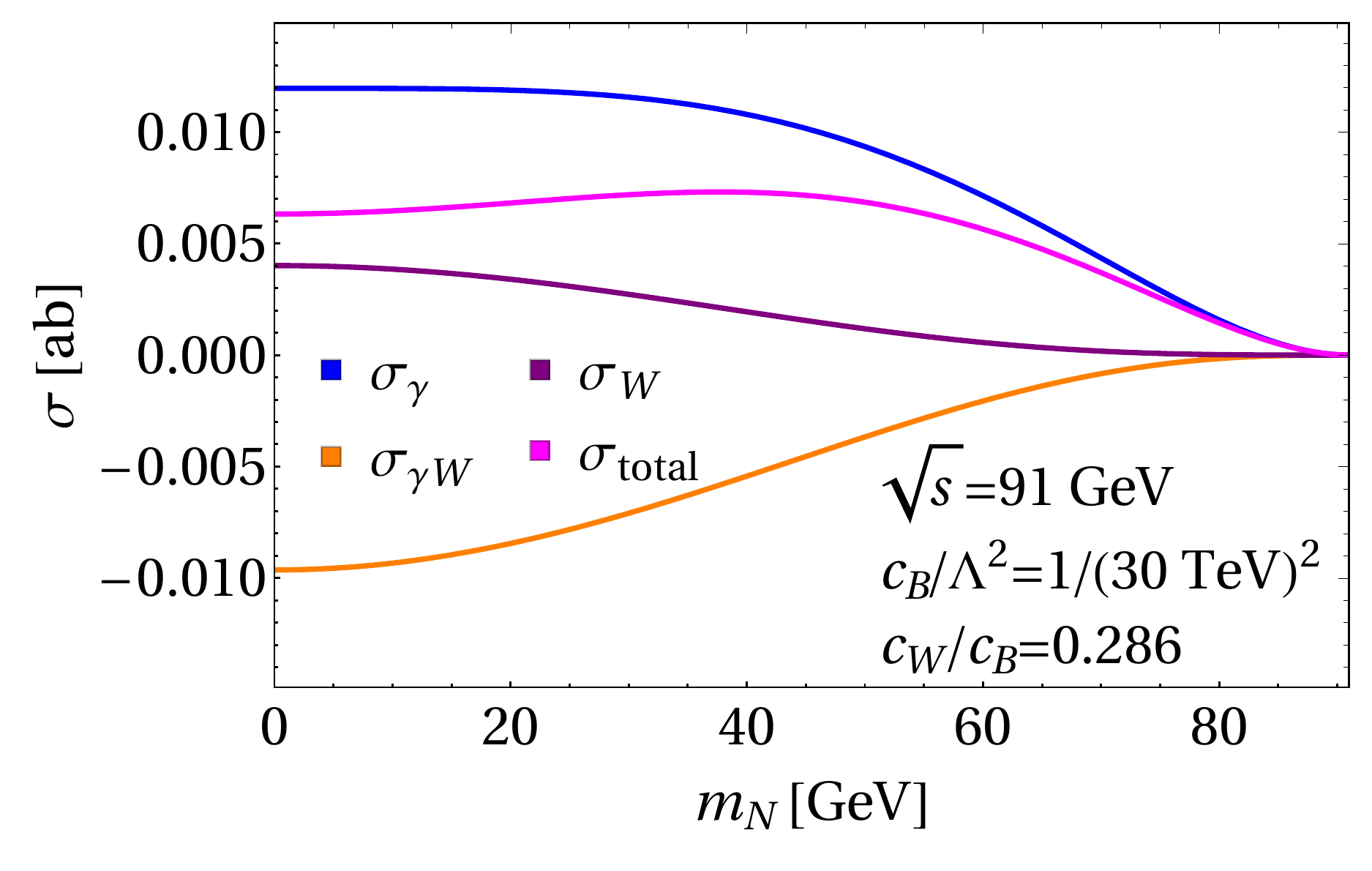}  &
    \includegraphics[width=0.48\textwidth]{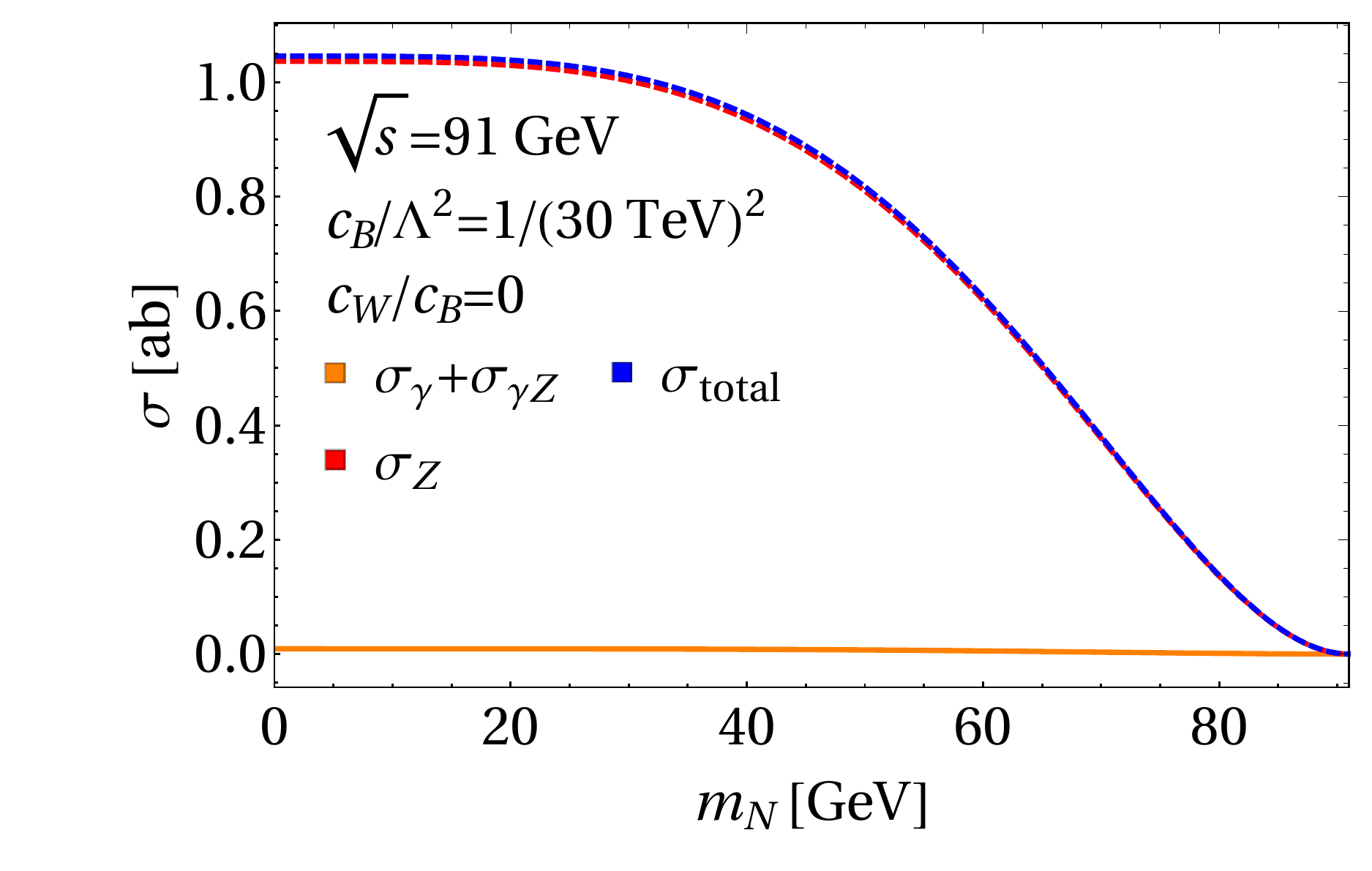}  \\ 
  \end{tabular}
 \caption{Cross section for sterile neutrino production via 2-2 process ($\ell^+ \ell^- \rightarrow \nu N$) with $c_B/\Lambda^2=1/(30 \text{ TeV})^{2}$ at $e^+e^-$ collider. \newtext{Cross section notations are described in \cref{app:xsec}.} The center-of-mass energy for collisions is fixed to $\sqrt{s}=91\,\text{GeV}$ in order to match LEP. The left and right panels correspond to $d_Z = 0\,(c_W/c_B=0.286)$  and $d_W=0 \,(c_W=0)$ benchmark points, respectively.}
 \label{fig:xsecBRplot}
\end{figure}

Our limit on $d_\gamma^e$ for the $d_Z = 0$ benchmark point is comparable to the one shown in the Figure 9 of \cite{Magill:2018jla}, where $d_W = 0$ and $d_Z=0$ are taken. In a consistent effective field theory framework, these two parameters cannot be set to zero simultaneously while allowing for a nonzero $d_\gamma$. The reason why the limits are still comparable is because the diagrams involving $W$ boson  yield subleading contributions to the total cross section, see left panel in \cref{fig:xsecBRplot}. For completeness, we also show sterile neutrino production cross section for $d_W = 0$ benchmark point in the right panel of \cref{fig:xsecBRplot}. For $d_W = 0$, the constraint in \cref{fig:LEP91GeV} improves compared to the $d_Z = 0$ case due to the larger 2-2 production cross section, as evident by comparing the panels in \cref{fig:xsecBRplot}. This difference is governed by the absence of sterile neutrino production via the Z resonance for $d_Z=0$. 

We also employed the results from \cite{L3:1997exg}, which provide branching ratio limits for $Z \to \gamma X$, to calculate the constraint for the $d_W=0$ benchmark scenario. We found that $Z$ decays yield comparable limits to those from the above analysis, namely $d_\gamma \simeq 10^{-5} \text{ GeV}^{-1}$.

\par

\subsection{CEPC and FCC-ee}
\label{subsec:CEPC_FCCee}

\noindent
CEPC and FCC-ee are proposed electron-positron colliders with comparable center-of-mass energy and luminosity (see \cref{tab:exp_CMenergy_lumi}); hence, the sensitivity reach we present in this section applies to both. As for LEP, we consider only 0-$\gamma$ and $\gamma$-$\gamma$ channels. We do not consider channels with heavy gauge bosons, noting that at $\sqrt{s}=240$ GeV, the sterile neutrino production is phase space suppressed for $m_N \gtrsim 100$ GeV. 

\begin{figure}[h!]
  \centering
    \includegraphics[width=0.55\linewidth]{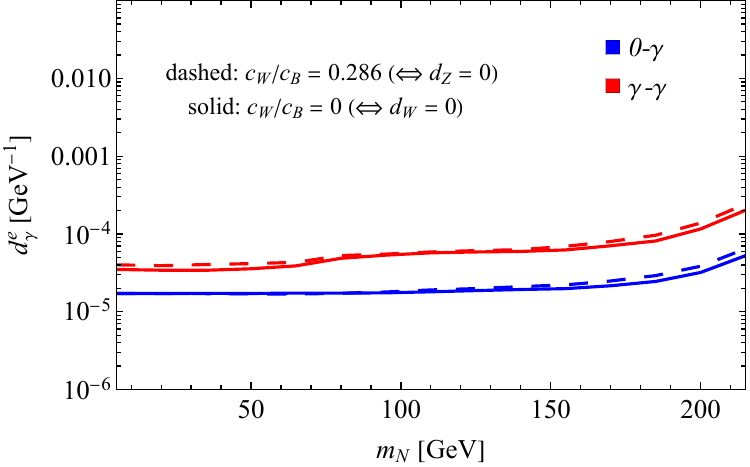} 
    \caption{Expected sensitivity for the $d_Z=0$ and $d_W=0$ benchmark points at CEPC and FCC-ee with $\sqrt{s}=240$ GeV.}
    \label{fig:CEPC240GeV}
\end{figure}

The generator-level and analysis cuts presented in \cref{sec:Event cuts,tab:CEPC240} were applied both to the signal and background events. We obtained \cref{fig:CEPC240GeV}, where we show expected sensitivity for $d_Z=0$ and $d_W=0$ benchmark points. As can be seen from that figure, the sensitivity for the $\gamma$-$\gamma$ channel (red) is comparable, but clearly subdominant across all $m_N$ values, to the 0-$\gamma$ channel (blue), given the larger cross section for 0-$\gamma$  channel at $\sqrt{s} \simeq \mathcal{O}(100)$ GeV (see \cref{fig:xsecBRplot2}).

We observe from \cref{fig:CEPC240GeV} that the expected sensitivity at CEPC/FCC-ee for $d_W = 0$ benchmark point is $d_\gamma^e\simeq 10^{-5}$ GeV$^{-1}$ which is similar to the LEP limit shown in \cref{fig:LEP91GeV}. At first glance, this appears surprising, given that CEPC/FCC-ee are expected to achieve a luminosity four orders of magnitude greater than LEP.
However, note that the 0-$\gamma$ signal cross section at LEP is approximately 100 times larger, given the sterile neutrino production via on-shell $Z$ boson. 
This can be observed by comparing the $\sigma$ values (blue lines) at $c_W/c_B=0$ between panels (a) and (b) of \cref{fig:xsecBRplot2}. The difference in luminosity and cross section compensates, explaining the similar results for the expected sensitivity at CEPC/FCC-ee and previously derived LEP limit. Despite that, note that CEPC/FCC-ee can probe higher values of $m_N$ because of the larger center-of-mass energy. As far as $d_Z = 0$ benchmark point is concerned, we show in \cref{fig:CEPC240GeV} almost identical expected sensitivity to the one for $d_W = 0$. This is because 2-2 sterile neutrino production cross sections for these two benchmark points are comparable for a fixed value of $c_B$.
Note that the result for $d_Z = 0$ presents a significant improvement over LEP, where for this benchmark point we found a limit of $d_\gamma^e \simeq 10^{-4}$ GeV$^{-1}$.
Without the absence of sterile neutrino production via the Z resonance at LEP, for $d_Z=0$, we observe a sterile neutrino production cross section at CEPC/FCC-ee comparable to that at LEP, with the higher CEPC/FCC-ee luminosity leading to improved results.

\subsection{MuC and CLIC}
\label{subsec:muonC}
\noindent
In addition to $\sqrt{s}\simeq \mathcal{O}(100)$ GeV colliders, scrutinized in \cref{subsec:LEP,subsec:CEPC_FCCee}, we also consider the proposed TeV-scale lepton colliders. We mainly focus on the MuC, that has been discussed in the context of two different center-of-mass energies, $\sqrt{s}=3 \text{ TeV}$ and $10 \text{ TeV}$. The integrated luminosities for these two setups are expected to be  $3\text{ ab}^{-1}$ and  $10\text{ ab}^{-1}$, respectively, following from $\mathcal{L} = 10 \text{ ab}^{-1} \big({\sqrt{s}}/{10\text{ TeV}}\big)$ \cite{Accettura:2023ked}. 

The envisioned maximum center-of-mass energy and integrated luminosity for the proposed electron-positron collider CLIC is very similar to that of MuC at 3 TeV \cite{Brunner:2022usy}. Consequently, both MuC at $\sqrt{s}=3$ TeV and CLIC are expected to yield comparable sensitivities for the transition magnetic moment. 
There is, however, a caveat: MuC and CLIC involve different colliding leptons ($\mu^+ \mu^-$ and $e^+ e^-$, respectively), and the sensitivities derived in this work apply strictly to the scenario where the sterile neutrino interacts with neutrino and charged lepton flavor matching the flavor of the colliding leptons. Indeed, examining \emph{e.g.} the left and middle diagrams in \cref{fig:diag2}, one can infer that certain diagrams are present only if the flavor of the colliding leptons matches that of the sterile neutrino interaction. This implies that the expected sensitivities for $d_\gamma^\mu$ at a 3 TeV MuC would also apply to CLIC, but for $d_\gamma^e$. With this in mind, the following discussion on the 3 TeV MuC also applies to CLIC.

\par
\begin{figure}[h!]
  \centering
  \begin{tabular}{cc}
    $\quad \quad $ (a) & $\quad \quad$  (b) \\
    \includegraphics[width=0.45\textwidth]{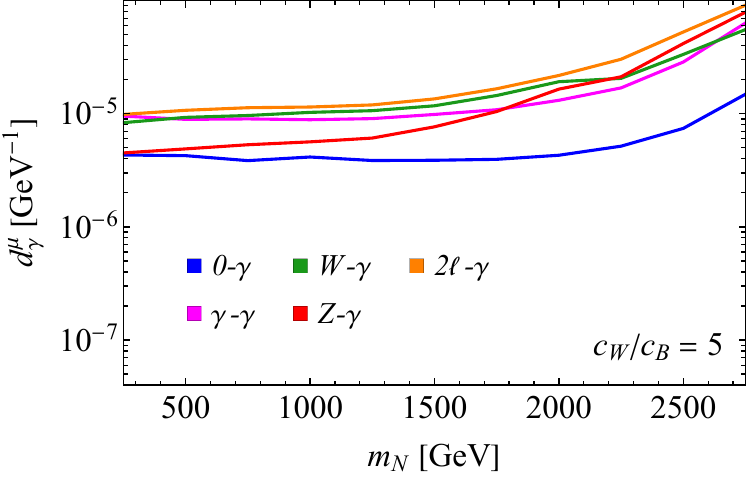} 
    \label{fig:muC3Tcwcb5gamma} &
    \includegraphics[width=0.45\textwidth]{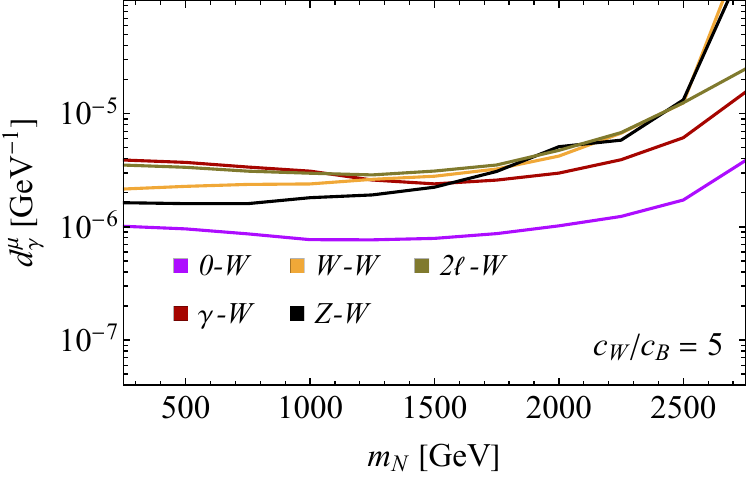} 
    \label{fig:muC3Tcwcb0w}\\ 
  \end{tabular}

  \begin{tabular}{c}
       $\quad  \quad$ (c)  \\
    \includegraphics[width=0.45\textwidth]{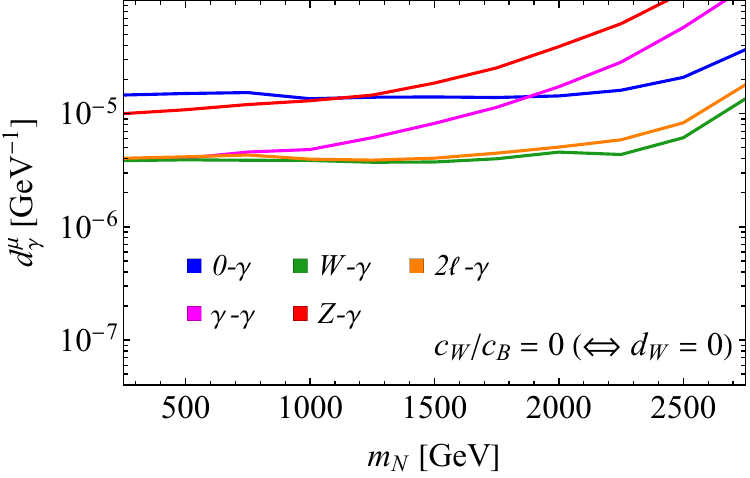}
    \label{fig:muC10Tcbcw3.27901}  \\
  \end{tabular}
    \caption{Expected sensitivity for $d_\gamma^\mu$ at the $\sqrt{s}=3$ TeV MuC. The benchmark point
    $c_W/c_B=5$ ($d_W=0$) is considered in upper (lower) panel(s). The results also hold for $d_\gamma^e$ at CLIC.}  
     \label{fig:muonC3TeV}
\end{figure}

\par
\begin{figure}[h!]
  \centering
  \begin{tabular}{cc}
    $\quad \quad \quad \quad$ (a) & $\quad \quad \quad \quad$  (b) \\
    \includegraphics[width=0.45\textwidth]{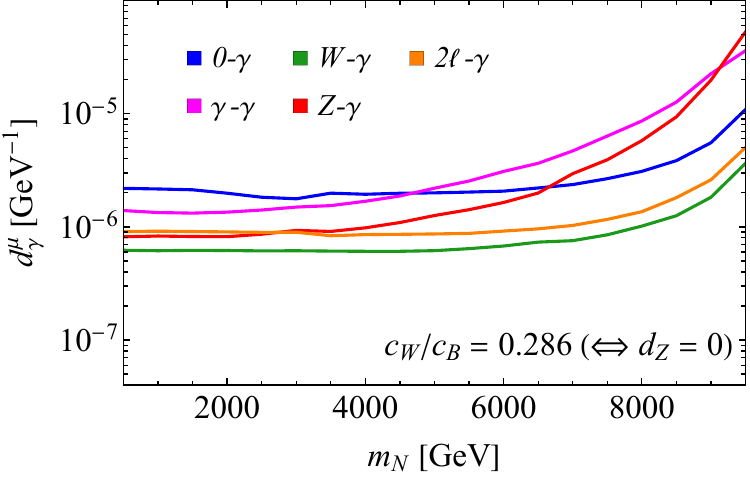} 
    \label{fig:muC10Tcbcw3.27901gamma} &
    \includegraphics[width=0.45\textwidth]{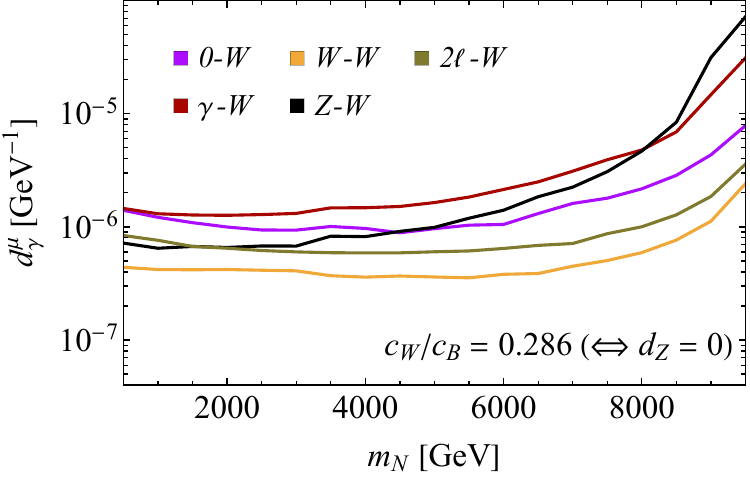} 
    \label{fig:muC10Tcbcw3.27901w}\\ 
    & \\ 
       $\quad \quad \quad \quad$ (c) & $\quad \quad \quad \quad$  (d) \\
    \includegraphics[width=0.45\textwidth]{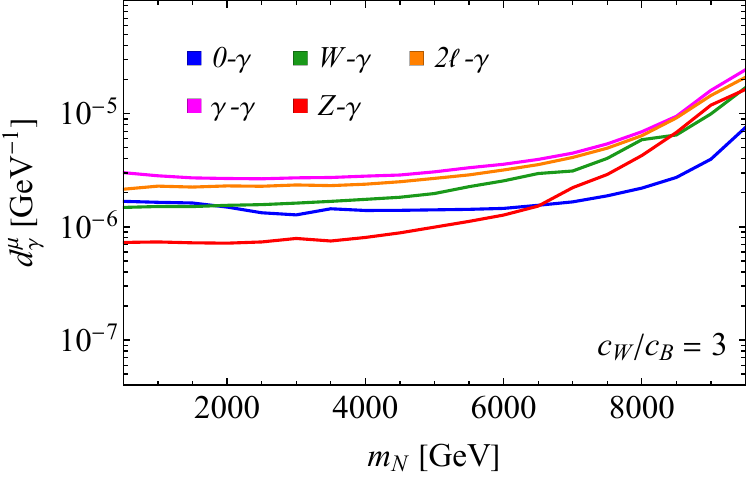}
    \label{fig:muC10Tcwcb3gamma} &
    \includegraphics[width=0.45\textwidth]{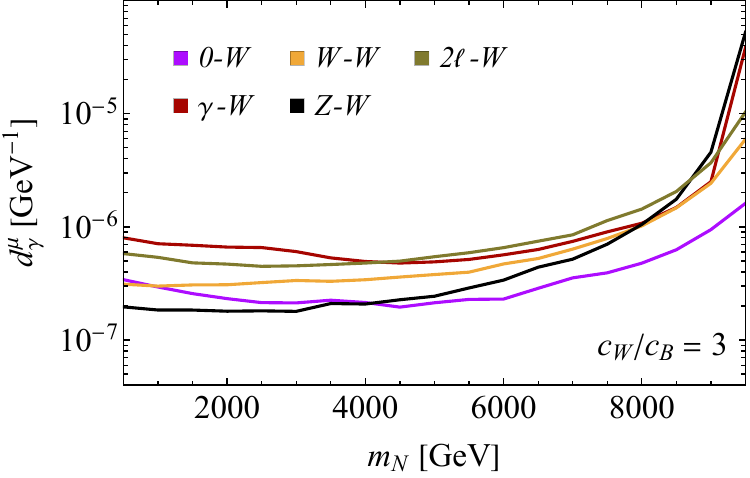}
    \label{fig:muC10Tcwcb3w} \\ 
  \end{tabular}
    \caption{Expected sensitivity for $d_\gamma^\mu$ at the $\sqrt{s}=10$ TeV MuC. The benchmark point $d_Z=0$ ($c_W/c_B=3$) is considered in upper (lower) panels. }    \label{fig:muonC10TeV}
\end{figure}

For MuC we consider all 10 interaction channels listed in \cref{tab:all_channels}. For particular channels, the results strongly depend on $c_W/c_B$ and therefore we choose to study several benchmark points in order to illustrate such behavior. Specifically, we consider $(i)$ $c_W/c_B=5$  and $(ii)$ $d_W=0$ for $\sqrt{s}=3$ TeV MuC and $(iii)$ $d_Z=0$ and $(iv)$ $c_W/c_B=3$ for $\sqrt{s}=10$ TeV MuC. The generator-level and analysis cuts presented in \cref{sec:Event cuts,tab:muonC3,tab:muonC3part2,tab:muonC10,tab:muonC10part2} were applied both to the signal and background events. In \cref{fig:muonC3TeV,fig:muonC10TeV} we present expected sensitivities for $d_\gamma^\mu$ at 3 TeV and 10 TeV MuC, respectively, and in what follows we discuss these results in detail. 

In the upper panels of \cref{fig:muonC3TeV} we consider benchmark point $c_W/c_B=5$ and we separate scenarios where sterile neutrino decays via $N\to\nu\gamma$ (left panel) and $N\to \ell^\pm W^\mp$ (right panel). Looking at both panels separately, one can observe that 0-$\gamma$ and 0-$W$ channels, for which sterile neutrinos are produced via 2-2 process, yield the strongest sensitivity. For this benchmark point at 3 TeV MuC, the 2-2 sterile neutrino production cross section is indeed largest (see panel (c) in \cref{fig:xsecBRplot2}). Note, however, that 0-$\gamma$ process still yields weaker sensitivity compared to several other channels in the right panel, in addition to  
0-$W$. This can be understood from the larger branching ratio for $N\to \ell^\pm W^\mp$ (see panel (d) in \cref{fig:xsecBRplot3}) as well as from the invariant mass cut. Regarding latter, if sterile neutrino decays to a $W$ boson and a charged lepton, its mass can be reconstructed and this serves as a strong signal-to-background discriminator, see details in \cref{sec:Event cuts}.

According to \cref{fig:xsecBRplot2}, the 2-3 and 2-4 sterile neutrino production processes become dominant as $|c_W/c_B|$ decreases and therefore in panel (c) of \cref{fig:muonC3TeV}, where we show results for $d_W=0$ benchmark point, the situation is very different compared to $c_W/c_B=5$ case. First, note that since the coupling of the sterile neutrino to a $W$ boson vanishes at tree level for this benchmark point, $N$ primarily decays into photons. Therefore, we consider only one decay channel for this benchmark, shown in panel (c) of \cref{fig:muonC3TeV}.
We find $W$-$\gamma$ and $2\ell$-$\gamma$ channels to yield overall the strongest sensitivity, in agreement with the expectation from comparing the cross sections between different production channels. While we focus on $d_\gamma^\mu$, we note that $2\ell$-$\gamma$ channel yields sensitivity also for $d_\gamma^e$ and $d_\gamma^\tau$ scenarios, as can be inferred from the topology of the diagram shown in the right panel of \cref{fig:diag2}. Another thing to note in connection to panel (c) is that, unlike for $c_W/c_B=5$ benchmark point as well as colliders studied in \cref{subsec:LEP,subsec:CEPC_FCCee}, we find that for $m_N \lesssim 2$ TeV $\gamma$-$\gamma$ yields stronger sensitivity compared to the vanilla 0-$\gamma$ channel. This again follows from the larger cross section for sterile neutrino production via 2-3  process in comparison to 2-2. The 2-3 cross section, however, falls more rapidly at larger $m_N$, allowing sensitivity for 0-$\gamma$ to surpass $\gamma$-$\gamma$ for $m_N \simeq \sqrt{s}$. The more rapid fall of the cross section for $d_W = 0$ occurs because the diagram shown in the middle panel of \cref{fig:diag2} vanishes at tree level; the other diagrams for the 2-3 $\gamma$-$\gamma$ channel, which lead to a more forward $\gamma/Z$ in the off-shell regime, are further suppressed as $m_N$ increases.

Expected sensitivities for $\sqrt{s}=10$ TeV MuC are presented in \cref{fig:muonC10TeV} for $d_Z=0$ (upper panels) and $c_W/c_B=3$ (lower panels) benchmark points. The $d_Z=0$ case corresponds to $c_W/c_B\simeq 0.286$ and we observe from (d) panel in \cref{fig:xsecBRplot2} that in this scenario all considered 2-3 processes have larger cross section compared to the 2-2 one. In accord with that, in panels (a) and (b) we find the strongest expected sensitivity for $W$-$\gamma$ and $W$-$W$ channels, respectively. Among these two, $W$-$W$ channel can be used for probing smaller values of $d_\gamma^\mu$, reaching $d_\gamma^\mu \simeq 5\times 10^{-7}$ GeV$^{-1}$. Note that, unlike for $c_W/d_B=5$ discussed above, 
the $N\to \ell^\pm W^\mp$ decay rate is smaller than $N\to\nu\gamma$ one, preventing the sensitivities in the right panel from strongly surpassing those on the left panel, despite the employed invariant mass cuts.

In contrast, for $c_W/c_B=3$ benchmark point, shown in the lower panels of \cref{fig:muonC10TeV}, the expected sensitivity for channels where sterile neutrino decays to a $W$ boson is observably better, which, again, follows from a larger $N\to \ell^\pm W^\mp$ decay branching ratio and the invariant mass cut. Let us demonstrate that by quantitatively comparing  $Z$-$\gamma$ and $Z$-$W$ channels from panels (c) and (d), respectively. These two channels yield the strongest sensitivities in the respective panels for low values of $m_N$. For these two, sterile neutrino production cross section is identical ($\mu^+ \mu^- \to N Z \nu$) and $N\to \ell^\pm W^\mp$   partial decay width is $\simeq 3$ times larger than $N\to\nu\gamma$ one. At $m_N \simeq$ 1 TeV, the sensitivity associated to the $Z$-$W$ channel is roughly 4 times stronger than the one corresponding to $Z$-$\gamma$, reaching $d_\gamma^\mu \simeq 2 \times 10^{-7}$ GeV$^{-1}$. Given that the signal cross section scales as $\sigma_S\propto (d_\gamma^\mu)^2$, the efficiency improvement stemming from the invariant mass cut should be around $\simeq 5$, which is what we confirmed by separately studying the impact of that cut. Finally, we note that the sensitivity for 0-$W$ channel in panel (d) starts to dominate at larger $m_N$; this is because, as $m_N$ increases, 2-2 cross section, dominated by the t-channel process, falls less steeply compared to the cross sections corresponding to 2-3 processes.

Notice that in connection to \cref{fig:muonC3TeV,fig:muonC10TeV} we discussed several different channels for probing $d_\gamma^\mu$.  Specifically, for $c_W/c_B=5$ we found $0$-$W$ channel to yield the strongest sensitivity while for $d_W=0$ benchmark point $W$-$\gamma$ and $2\ell$-$\gamma$ are leading at $\sqrt{s}=3$ TeV. For $\sqrt{s}=10$ TeV, we considered $d_Z=0$ for which $W$-$W$ channel shows the highest sensitivity, while for $c_W/c_B=3$ we found $Z$-$W$ and 0-$W$ to be the most sensitive.

Finally, let us bring up recent study exploring transition magnetic moment at MuC \cite{Barducci:2024kig}. In that work, the authors introduced 2-3 sterile neutrino production process at MuC. In this section, we complemented and extended this picture by including more channels and benchmark scenarios. Specifically, the authors of \cite{Barducci:2024kig} considered 4 out of 10 channels listed in \cref{tab:all_channels}. These 6 additional channels are as follows: for the sterile neutrino production we have $\mu^+ \mu^- \rightarrow \nu Z (\gamma) N$ as well as vector boson fusion process ($\mu^+ \mu^- \rightarrow \nu  \mu^+ \mu^- N$). For each of the three aforementioned processes we consider sterile neutrino decay to both $\gamma$ and $W$, making it 6 novel channels in total.  

The cuts across the two analyses are also different; importantly, we utilize invariant mass cut that strongly improves the sensitivity for channels featuring
$N\to \mu^\pm W^\mp$ decay. Further, for muons, the authors of \cite{Barducci:2024kig} selected $\eta_{\mu}<7$ cut, possibly with the strategy of relying on the existence of a forward muon detector \cite{Ruhdorfer:2023uea}. In our work, we find comparable results for channels arising from the 2-2 sterile neutrino production process, assuming only the existence of a central detector and implementing the invariant mass cut.

We have, however, identified a discrepancy in the cross sections for 2-3 processes between our analysis and that of \cite{Barducci:2024kig}, which propagates into differences in the expected sensitivities. Specifically, we find these cross sections to be smaller, leading to more conservative results for the accessible active-to-sterile neutrino transition magnetic moment. To investigate this further, we analytically computed squared amplitudes and performed a 3-body phase space integration to obtain the 2-3 cross section for several benchmark scenarios. We found excellent agreement with results from \texttt{MadGraph5\_aMC@NLO}, 
strengthening the results of our analysis.

\section{Conclusions}
\label{sec:conclusions}
\noindent
In this work, we studied a model featuring a sterile neutrino, $N$, interacting with the Standard Model via the active-to-sterile neutrino magnetic moment $d_\gamma^\alpha$. We focused on exploring the potential of proposed future lepton colliders, including CEPC, FCC-ee, CLIC, and two MuC realizations (3 TeV and 10 TeV), to probe this interaction. Additionally, we revisited constraints from LEP. In our analysis, several sterile neutrino production processes were considered, including 2-2 process $\ell^+ \ell^- \to \nu N$, various 2-3 processes involving $\gamma/Z$-$\nu$ and $W \ell$ fusion and 2-4 vector boson fusion process. By incorporating the two primary decay modes of $N$ ($N\to \nu\, \gamma$ and $N\to \ell\, W$), we identified six channels with sizable cross sections for MuC and CLIC, along with the $\gamma$-$\gamma$ channel at CEPC, FCC-ee, and LEP, none of which have been previously explored in the literature.  
For all considered processes, we implemented the generator-level and analysis cuts and performed cut-and-count analysis in order to calculate the expected sensitivities.

Across several benchmark scenarios, we observed that the expected sensitivities vary significantly depending on sterile neutrino production channel, center-of-mass energy, and the flavor of neutrino and charged lepton interacting with the sterile neutrino. For LEP, CEPC and FCC-ee, the dominant channel involves sterile neutrino production via $e^+ e^- \to \nu N$, followed by its decay into $\gamma \nu$. In contrast, for MuC and CLIC, the situation is more complex, with multiple channels contributing. Among the 10 channels listed in \cref{tab:all_channels}, five--namely 
$0$-$W$, $W$-$\gamma$, $2\ell$-$\gamma$, $W$-$W$, and $Z$-$W$--emerged as the most sensitive in specific regions of the parameter space. Notably, for the studied benchmark points, values of $d_\gamma \simeq 10^{-7} \, \text{GeV}^{-1}$ are within reach at MuC and CLIC, significantly surpassing existing constraints for $m_N \gtrsim 100$ GeV. Our derived sensitivity $d_\gamma \simeq 10^{-7} \, \text{GeV}^{-1}$ covers the region of a dipole operator estimated with a cutoff scale $\Lambda = 20~\text{TeV}$ (above the collision energy of a $10~\text{TeV}$ MuC) and Wilson coefficients of $\mathcal{O}(1)$. This demonstrates that MuC can provide efficient probes of UV theories at this scale in the strong coupling regime.

\begin{acknowledgments}
\noindent
We would like to thank Daniele Barducci, Natascia Vignaroli, Akanksha Bhardwaj, and Ian M. Lewis for helpful discussions. The work of VB is supported by the United States Department of Energy Grant No. DE-SC0025477. VB would like to thank the Center for Theoretical Underground Physics and Related Areas (CETUP*) and the Institute for Underground Science at Sanford Underground Research Facility (SURF) for providing a conducive environment during the 2024 summer workshop. This work was performed in part at Aspen Center for Physics, which is supported by National Science Foundation grant PHY-2210452. 
\end{acknowledgments}

\bibliographystyle{JHEP}
\bibliography{refs}
\newpage

\appendix
\section*{Appendix}

\section{cross section for the 2-2 process}
\label{app:xsec}
\noindent
We present here the results for the case where the neutrino flavor produced in the s-channel matches the one from the t-channel, allowing for the interference across all diagrams. The cross section for the sterile neutrino production at lepton colliders
through 2-2 process, shown in \cref{fig:diag1}, consists of several components
\begin{align}
    \sigma_{tot} =  \sigma_{\gamma} + \sigma_{Z} + \sigma_{W} + \sigma_{\gamma Z} + \sigma_{\gamma W} + \sigma_{W Z}\,.
    \label{eq:sigtot}
\end{align}
Here, $\sigma_{\gamma}$, $\sigma_{Z}$ and $\sigma_{W}$ represent the contribution from the diagram including photon propagator, $Z$ and $W$ propagator, respectively and the terms with 2 indices arise from the interference between the diagrams involving different propagators. We obtained the following expressions 
\begin{eqnarray}
    \sigma_{\gamma} & = & \frac{\alpha  d_{\gamma}^2 \left(m_N^2-s\right)^2 \left(2 m_N^2+s\right)}{3 s^3}\,, \nonumber\\
 \sigma_{\gamma Z} & = & \frac{\alpha  d_{\gamma} d_Z \left(c_w^2-3 s_w^2\right) \left(m_N^2-s\right)^2 \left(2
   m_N^2+s\right)\left(s-m_Z^2\right)}{6 c_w s_w s^2 \left[\left(s-m_Z^2\right)^2+m_Z^2\Gamma_Z^2\right]}\, \nonumber\\
 \sigma_{\gamma W} & = & \frac{\alpha  d_{\gamma} d_W}{\sqrt{2} s_w s^2}  \Bigg[2 m_W^2 \left(-m_N^2+s+m_W^2\right) \log
   \frac{-m_N^2+s+m_W^2}{m_W^2} - \left(m_N^2-s\right) \left(m_N^2-s-2 m_W^2\right)\Bigg]\,,\nonumber\\
    \sigma_{W} & = & -\frac{\alpha  d_W^2 \left(m_N^2-s\right)}{2 s_w^2 s^2} \left[2 \left(m_N^2-s\right) - \left(m_N^2-s-2 M_W^2\right)\log
   \frac{-m_N^2+s+m_W^2}{m_W^2}\right]\,, \nonumber\\
 \sigma_{WZ} & = & -\frac{\alpha  d_W d_Z 
     \left(c_w^2-s_w^2\right)\left(s-m_Z^2\right)}{2 \sqrt{2} c_w s_w^2 s \left[\left(s-M_Z^2\right)^2+m_Z^2\Gamma_Z^2\right]}\times
\nonumber\\ & &     
      \Bigg[\left(m_N^2-s\right)
   \left(m_N^2-s-2 m_W^2\right) - 2 m_W^2 \left(-m_N^2+s+m_W^2\right) \log
   \frac{-m_N^2+s+m_W^2}{m_W^2}\Bigg]\,, \nonumber\\
 \sigma_{Z} & = & \frac{\alpha  d_Z^2 \left(c_w^4-2 c_w^2 s_w^2+5 s_w^4\right)
   \left(m_N^2-s\right)^2 \left(2 m_N^2+s\right) }{24 c_w^2 s_w^2 s \left[\left(s-m_Z^2\right)^2+m_Z\Gamma_Z^2\right]}\,,
   \label{eq:xsec}
\end{eqnarray}
where $m_N$ is the sterile neutrino mass, $s$ is the Mandelstam variable and, for brevity, we defined $c_w\equiv \cos \theta_w$ and
$s_w\equiv \sin \theta_w$. In contrast to previous studies \cite{Zhang:2023nxy}, we find non-vanishing contributions to the cross section from $\sigma_{\gamma W}$ and $\sigma_{W Z}$ terms. 

\par
\newpage
\section{List of cuts}
\label{sec:Event cuts}
\noindent
We define our data sample to meet the following conditions at the generator level:
\begin{itemize}
    \item \textbf{Photon}
    \begin{itemize}
     \item \textbf{LEP}:  $p_{T,\gamma}>3$ GeV, ($45^0<\theta_\gamma<135^0$ for 0-$\gamma$ channel and $|\eta|<2.5$ for $\gamma$-$\gamma$ channel). For 0-$\gamma$, the requirements on $\theta_\gamma$, the polar angle between the photon direction and the beam axis, are taken from \cite{DELPHI:1996drf}\,; \\
     \textbf{CEPC}: $p_{T,\gamma}>5$ GeV and $|\eta_\gamma| < 2.5$\,; \\
     \textbf{MuC}: $p_{T,\gamma}>10$ GeV and $|\eta_\gamma| < 2.5$\,. 
     \item require appropriate number of photons in the event for a given channel (see again \cref{tab:all_channels}): 
     \begin{enumerate}
         \item $n_\gamma=0$ (0-$W$, $W$-$W$, $Z$-$W$, $2\ell$-$W$)\,;
         \item $n_\gamma=1$ (0-$\gamma$, $W$-$\gamma$, $Z$-$\gamma$, $2\ell$-$\gamma$, $\gamma$-$W$)\,;
         \item $n_\gamma=2$ ($\gamma$-$\gamma$)\,.
     \end{enumerate} 
    \end{itemize}

    \item \textbf{Lepton}
    \begin{itemize}
    \item  $p_{T,\ell}>10$ GeV and $|\eta_\ell| < 2.5$\,.
        \item require appropriate number of leptons in the event for a given channel: 
         \begin{enumerate}
         \item $n_\ell=0$ (0-$\gamma$, $\gamma$-$\gamma$, $Z$-$\gamma$)\,;
         \item $n_\ell=1$ ($W$-$\gamma$, 0-$W$, $\gamma$-$W$, $Z$-$W$)\,;
         \item $n_\ell=2$ ($W$-$W$, $2\ell$-$\gamma$)\,;
         \item $n_\ell=3$ ($2\ell$-$W$)\,.
         \end{enumerate} 
    \end{itemize}
\end{itemize}

The above kinematics cuts are the generator-level cuts. For specific channels, we  apply stronger generator-level cuts in order to improve the efficiencies for generating signal and background in the desired phase space. Those channels and cuts are listed in \cref{tab:stronger_gen_cut_muonC}. For channels with multiple charged leptons,
$p^{(1)}_{T,\ell}$ denotes the transverse momentum of the highest-$p_T$ lepton.

\par
\begin{table}[h!]
    \centering
    \begin{tabular}{|c|c|c|}
    \hline
    Label &  MuC-3  & MuC-10  \\
    \hline
    \hline 
    \multirow{2}{*}{$\gamma$-$W$} & $p_{T,\ell}>500$ GeV, $p_{T,W}>200$ GeV & $p_{T,\ell}>1500$ GeV, $p_{T,W}>500$ GeV\\
    &                       $|\eta_W|<2$, $|\eta_\ell|<2.5$, $|\eta_\gamma|<2.5$  &  $|\eta_\ell|<2.5$, $|\eta_\gamma|<2.5$  \\
    \cline{1-3}
    \multirow{2}{*}{$Z$-$W$}  & $p_{T,\ell}>400$ GeV, $p_{T,W}>200$ GeV & $p_{T,\ell}>1300$ GeV, $p_{T,W}>600$ GeV\\
    &                       $|\eta_Z|<1$, $|\eta_\ell|<2.5$ & $|\eta_Z|<1$, $|\eta_\ell|<2.5$ \\
    \cline{1-3}
    \multirow{2}{*}{$W$-$W$}  & $p_{T,W}>200$ GeV, $p^{(1)}_{T,\ell}>500$ GeV & $p_{T,W}>200$ GeV, $p^{(1)}_{T,\ell}>1700$ GeV\\
    &                       $|\eta_W|<1$, $|\eta_\ell|<2.5$ & $|\eta_W|<1$, $|\eta_\ell|<2.5$ \\
    \cline{1-3}
    \multirow{2}{*}{$2\ell$-$W$}  & $p_{T,W}>300$ GeV, $p^{(1)}_{T,\ell}>600$ GeV & $p_{T,W}>700$ GeV, $p^{(1)}_{T,\ell}>1700$ GeV\\
    &                       $|\eta_\ell|<2.5$ & $|\eta_\ell|<2.5$ \\
    \hline
    \end{tabular}
    \caption{The generator-level cuts for specific channels.}
    \label{tab:stronger_gen_cut_muonC}
\end{table}

In \cref{tab:bg_xsec_list_muonC,tab:bg_xsec_list_LEPCEPC}, we show the background cross sections after imposing the generator-level cuts discussed above.

\par
\begin{table}[h!]
    \centering
    \begin{tabular}{|c|c|c|c|}
    \hline
    Label & Background & MuC-3 ($\sqrt{s}=3$ TeV) (pb)& MuC-10 ($\sqrt{s}=10$ TeV) (pb)\\
    \hline
    \hline
     0-$\gamma$ & $\mu^+ \mu^- \rightarrow \gamma + \nu\nu/\nu\nu\nu\nu$ &  3.009&  3.294\\
    \hline
    $\gamma$-$\gamma$ & $\mu^+ \mu^- \rightarrow \nu \nu \gamma \gamma$ &  0.079&  0.095\\
    \hline
    $Z$-$\gamma$ & $\mu^+ \mu^- \rightarrow \nu Z \nu \gamma $ &  0.136&  0.276\\
    \hline
    $W$-$\gamma$ & $\mu^+ \mu^- \rightarrow \mu^{\pm} W^{\mp} \nu \gamma$ &  0.050&  0.020\\
    \hline
    $2\ell$-$\gamma$ & $\mu^+ \mu^- \rightarrow \nu_{l_1} \ell ^{\pm} _1 \nu_{l_2} \ell^{\mp} _2 \gamma $ &  0.021&  0.016\\
    \hline
    
    0-$W$ & $\mu^+ \mu^- \rightarrow W^\pm \ell^\mp \nu$ & 1.051 & 0.367 \\
    \hline
    $\gamma$-$W$ & $\mu^+ \mu^- \rightarrow \nu \gamma \mu^{\pm} W^{\mp} $ &  0.0092 &  0.0027\\
    \hline
    $Z$-$W$ & $\mu^+ \mu^- \rightarrow \nu Z \mu^{\pm} W^{\mp} $ &  0.0048 &  0.0016\\
    \hline
    $W$-$W$ & $\mu^+ \mu^- \rightarrow \mu^{\pm} W^{\pm} \mu^{\mp} W^{\mp}$ &  0.00082&  0.00042\\
    \hline
    $2\ell$-$W$ & $\mu^+ \mu^- \rightarrow  \ell_1 ^{\pm} \ell_1 ^{\mp} \ell^{\pm} _3 \nu  W^\mp $ & 0.00086  & 0.00039 \\
    \hline
    \end{tabular}
    \caption{The relevant background cross sections after generator-level cuts for MuC at two center-of-mass energies.}
    \label{tab:bg_xsec_list_muonC}
\end{table}
\begin{table}[h!]
    \centering
    \begin{tabular}{|c|c|c|c|}
    \hline
    Label & Background & LEP ($\sqrt{s}=91$ GeV) (pb)& CEPC/FCC-ee ($\sqrt{s}=240$ GeV) (pb)\\
    \hline
    \hline
     0-$\gamma$ & $\mu^+ \mu^- \rightarrow \gamma + \nu\nu/\nu\nu\nu\nu$ &  4.050 &  3.476\\
    \hline
    $\gamma$-$\gamma$ & $\mu^+ \mu^- \rightarrow \nu \nu \gamma \gamma$ &  0.0095 &  0.118\\
    \hline
    \end{tabular}
    \caption{The relevant background cross sections after generator-level cuts for LEP and CEPC/FCC-ee.}
    \label{tab:bg_xsec_list_LEPCEPC}
\end{table}

We impose further cuts at the analysis level. The most significant analysis cuts are as follows:
\begin{itemize}

    \item \emph{Veto Z-resonance}: Let $p_X$ be the sum of 4-momenta of all the visible particles with the invariant mass, $p_X ^2 = m_X ^2$. The resonance energy of the invisible particles (from on-shell $Z$ boson decays) reads
    \begin{equation}
        E_{Z} ^{\text{res}}= \frac{s + m_X ^2 - m_Z ^2}{2\sqrt{s}}\,.
        \label{eq:Z-resonance}
    \end{equation}
    We require $E_X \notin [E_{Z} ^{\text{res}} - 5 \, \Gamma_Z ^\text{boost}, E_{Z} ^{\text{res}} +  5 \, \Gamma_Z ^\text{boost}]$, where $ \Gamma_Z ^\text{boost} = m_Z \Gamma_Z /\sqrt{s}$ is the decay width of the boosted $Z$ boson and $\Gamma_Z=2.49 $ GeV.
    
    \item \emph{Missing transverse energy}:  This cut is applied to the channels with 
neutrino(s) in the final state. Both $E^{miss}_T > m_N/4$ and $E_T^{\text{miss}}>20$ GeV are imposed for the channels where neutrino is one of the decay products of $N$. We set $E_T^{\text{miss}}>20$ GeV when neutrino is not  produced from $N$ in order to avoid stringent cut and respective loss of signal events for high values of $m_N$.

    \item \emph{Reducible backgrounds}: The backgrounds ($\ell^+ \ell^- \rightarrow \ell^+ \ell^- \gamma$ and $\ell^+ \ell^- \rightarrow  \gamma\gamma\gamma$) to the 0-$\gamma$ channel can be removed by imposing boundary-dependent cut \cite{Liu:2019ogn}. We define the angle $\theta_B$ corresponding to the boundary of the electromagnetic calorimeter. We find the maximum allowed energy for the photon
    \begin{equation}
        E_\gamma^m(\theta_\gamma) = \sqrt{s} \left[ 1 + \frac{\sin \theta_\gamma}{\sin \theta_B} \right]^{-1}\,.
        \label{eq:dd}
    \end{equation}
    
    We set $| \cos \theta_B| = 0.99$ for the boundary and add the following cuts for the 0-$\gamma$ channel:
    \begin{itemize}
        \item single photon must be detected in the calorimeter, \emph{i.e.}, $\cos \theta_\gamma < \cos \theta_B$\,;
        \item photon energy must be greater than the value in \cref{eq:dd}, \emph{i.e.}, $E_\gamma (\theta_\gamma) > E_\gamma^m (\theta_\gamma)$\,.\\
        \end{itemize}

    \item \emph{Invariant mass $m_{W\ell}$}: This cut is applied for the channels where sterile neutrino decays into visible particles ($N \rightarrow W^\pm \ell^\mp $). We construct the invariant mass squared, $m_{W\ell} ^2 =(p_W+p_\ell)^2$, from a pair of $W$ boson and lepton $\ell$ with opposite charges. There can be multiple pairs of opposite charged $(W,\ell)$. We require one pair to satisfy $|m_{W\ell}-m_N|<20$ GeV. $W$ boson identification is discussed below. 
\end{itemize}

We do not explicitly decay $W$ and $Z$ bosons in our analysis, but when applying cuts to their kinematic properties, we need to ensure that only events in which all decay products can be reconstructed are considered. The $W$ boson decays either hadronically ($W\rightarrow jj$) or leptonically ($W\rightarrow\ell \nu$). The hadronic decay is favorable due to the larger branching ratio;  moreover, if $W$ decays leptonically, produced neutrino acts as a missing energy, preventing the full reconstruction of the parent particle when multiple neutrinos are involved. Similarly, hadronic decay mode of $Z$ boson ($Z\rightarrow jj$) is dominant, although decay into charged leptons ($Z\rightarrow \ell^+ \ell^-$) also leads to  the reconstruction of $Z$ boson. 

We modify the cross sections of the channels with $W$ and $Z$ bosons by including appropriate branching ratios and jet reconstruction efficiencies for the identification of these heavy vector bosons. The latter are set to 50\% \cite{ATLAS:2020szu} for both $W$ and $Z$ boson. Hence, we implement the efficiency for identifying $W$ and $Z$ bosons as  
\begin{align}
    r_W& =0.50 \times 0.6741 , \\ r_Z &=0.50 \times 0.6991 + 2\times 0.034, 
    \label{eq:reco_values}
\end{align}
where we used branching ratios for hadronic modes of $W$ (67.41\%) and $Z$ boson (69.91\%) as well as the branching ratios for leptonic modes of $Z$ boson  (3.4\% for electron pair or muon pair) \cite{ParticleDataGroup:2018ovx}. 
If the counts of produced $W$ and $Z$ bosons are $w$ and $z$, respectively, the modified cross section for a given channel $\mathcal{C}$ reads
\begin{equation}
       \sigma^{\text{tagged}}_\mathcal{C} = \big(r_W\big)^{w} \times \big(r_Z\big)^{z} \times  \sigma_\mathcal{C} \,. 
   \label{eq:xsec_wzboson_reco}
\end{equation}
The effective (tagged) cross section $\sigma^{\text{tagged}}_\mathcal{C}$ from \cref{eq:xsec_wzboson_reco} is used in \cref{eq:sig_bg_count} to calculate the number of signal and background events.\\

\par
\begin{table}[h!]
    \centering
    \begin{tabular}{|c|c|c|}
    \hline
    Cut & Signal & Background \\
    \hline
    \hline
    Veto Z-resonance & 1.0 & 0.79635 \\
    \hline
    Missing transverse energy & 0.9757 & 0.77758 \\
    \hline
    Invariant mass $m_{W\ell}$ & 0.9757 & 0.01142 \\
    \hline
    $(p_{T},\eta)_Z$ & 0.4679 & 0.00518 \\
    \hline
    $(p_{T},\eta)_W$ & 0.3468 & 0.00300 \\
    \hline
    $(p_{T},\eta)_\ell$ & 0.2607 & 0.00221 \\
    \hline
    All & 0.2607 & 0.00221 \\
    \hline
    \end{tabular}
    \caption{Cut flow table for the $Z$-$W$ channel at MuC ($\sqrt{s}=3$ TeV). For each cut, the fraction of signal and background removed is indicated. The benchmark point used for the signal is $c_W/c_B=5$ with $m_N=1500$ GeV.}
    \label{tab:cut_efficiency}
\end{table}

The remaining analysis cuts are shown separately for considered colliders in \cref{tab:LEP91,tab:CEPC240,tab:muonC3,tab:muonC3part2,tab:muonC10,tab:muonC10part2}.
There, we show the cuts associated to three $m_N$ regions (see main text for further discussion on $m_N$-dependent cuts). In these tables, when there is multiplicity in the particles (\emph{e.g.}, $\gamma$-$\gamma$, $W$-$W$, $2\ell$-$\gamma$, $2\ell$-$W$), we label them as $X_1,\,X_2$ such that the first particle has the largest $p_T$ ($p_{T,X_1} > p_{T,X_2}$). We do not apply these additional cuts in scenarios where they fail to provide sufficient improvement in distinguishing the signal from the background (\emph{e.g.}, $p_{T,\ell}$ for $W$-$\gamma$ channel in the second mass region, see \cref{tab:muonC3}). While we do not impose cuts on all visible particles, it is important to record their 4-momenta to construct $m_X$ in \cref{eq:Z-resonance}.

When appropriate, we also include in \cref{tab:LEP91,tab:CEPC240,tab:muonC3,tab:muonC3part2,tab:muonC10,tab:muonC10part2} the $m_N$-dependence of cuts in the high-$m_N$ region. When a particle is produced along with a sterile neutrino, it will have less energy available when $m_N$ is large. The maximum available energy is $\sqrt{s}-m_N$, which has been used as the maximum $p_T$ cut for some channels, \emph{e.g.}, $W$ boson in $W$-$W$ channel at $\sqrt{s}=3$ TeV.

Finally, as an example, we present  in \cref{tab:cut_efficiency} a cut flow table for the $Z$-$W$ channel. Notice that the invariant mass cut removes about 76\% of the background while keeping all the signal events. Such high efficiency yields strong improvement in the sensitivity for the channels featuring $N\rightarrow \mu W $ decay, as illustrated in \cref{fig:muonC3TeV,fig:muonC10TeV}.

\newpage

\section{Cut tables}
\label{sec:Cut tables}

\par
\begin{table}[h!]
    \caption{LEP (91 GeV) cuts.}
    \label{tab:LEP91} 
    \centering
    \begin{tabular}{|c|c|c|c|}
        \hline
        Process & Mass region (in GeV) & Particle ($x$) & $p_{T,x} >$ (GeV) ($<$ specified) \\
        \hline
        \hline
        \multirow{3}{*}{0-$\gamma$} 
            & \multirow{1}{*}{$m_N \leq 30$} & $\gamma$ & 15 \\ 
        \cline{2-4}
            & \multirow{1}{*}{$30<m_N \leq 60$} & $\gamma$ & 20 \\ 
        \cline{2-4}
            & \multirow{1}{*}{$60<m_N$} & $\gamma$ & 25 \\ 
        \hline
        \multirow{3}{*}{$\gamma$-$\gamma$} 
            & \multirow{1}{*}{$m_N \leq 30$} & $\gamma_1$ & 15 \\ 
        \cline{2-4}
            & \multirow{1}{*}{$30<m_N \leq 60$} & $\gamma_1$ & 20 \\ 
        \cline{2-4}
            & \multirow{1}{*}{$60<m_N$} & $\gamma_1$ & 25 \\ 
        \hline

    \end{tabular}
\end{table}

\par
\begin{table}[h!]
    \caption{CEPC and FCC-ee (240 GeV) cuts. }
    \label{tab:CEPC240} 
    \centering
    \begin{tabular}{|c|c|c|c|c|c|c|}
        \hline
        Process & Mass region (in GeV) & Particle ($x$) & $p_{T,x} >$ (GeV) ($<$ specified) & $|\eta|<$  \\
        \hline
        \hline
        \multirow{3}{*}{0-$\gamma$} 
            & \multirow{1}{*}{$m_N<80$} & $\gamma$ & 40 & 1  \\ 
        \cline{2-5}
            & \multirow{1}{*}{$80 \leq m_N<160$} & $\gamma$ & 50 & 1   \\
        \cline{2-5}
            & \multirow{1}{*}{$160 \leq m_N$} & $\gamma$ & 60 & 1  \\
        \hline
        \multirow{6}{*}{$\gamma$-$\gamma$} 
            & \multirow{2}{*}{$m_N \leq 80$} & $\gamma_1$ & 40 & 1 \\
            &                       & $\gamma_2$ & $-$ & 1 \\
        \cline{2-5}
            & \multirow{2}{*}{$80 \leq m_N<160$} & $\gamma_1$ & 60 & 1 \\
            &                       & $\gamma_2$ & $-$ & 1 \\
        \cline{2-5}
            & \multirow{2}{*}{$160 \leq m_N$} & $\gamma_1$ & 70 & 1 \\
            &                       & $\gamma_2$ & $-$ & 1 \\
        \hline

    \end{tabular}
\end{table}

\par
\begin{table}[h!]
    \caption{MuC (3 TeV) cuts.}
    \label{tab:muonC3}
    \centering
    \begin{tabular}{|c|c|c|c|c|}
        \hline
        Process & Mass region (in GeV) & Particle ($x$) & $p_{T,x} >$ (GeV) ($<$ specified) & $|\eta|<$ \\
        \hline
        \hline
        \multirow{3}{*}{0-$\gamma$} 
            & \multirow{1}{*}{$m_N \leq 750$} & $\gamma$ & 200 & $-$ \\ 
        \cline{2-5}
            & \multirow{1}{*}{$750<m_N<2250$} & $\gamma$ & 250 & 1 \\
        \cline{2-5}
            & \multirow{1}{*}{$2250 \leq m_N$} & $\gamma$ & 350 & 1 \\
        \hline
        \multirow{6}{*}{$\gamma$-$\gamma$} 
            & \multirow{2}{*}{$m_N \leq 750$} & $\gamma_1$ & 150 & 1.5 \\
            &                       & $\gamma_2$ & 50 & 1 \\
        \cline{2-5}
            & \multirow{2}{*}{$750<m_N<2250$} & $\gamma_1$ & 300 & 1 \\
            &                       & $\gamma_2$ & 50 & 1 \\
        \cline{2-5}
            & \multirow{2}{*}{$2250 \leq m_N$} & $\gamma_1$ & 400 & 1 \\
            &                       & $\gamma_2$ & 50 & 1 \\
        \hline
        \multirow{6}{*}{$Z$-$\gamma$} 
            & \multirow{2}{*}{$m_N \leq 750$} & $\gamma$ & 100 & $-0.5 < \eta < 1.5$ \\
            &                       & $Z$ & 300 & 1 \\
        \cline{2-5}
            & \multirow{2}{*}{$750<m_N<2250$} & $\gamma$ & 300 & 1 \\
            &                       & $Z$ & 300 & 1 \\
        \cline{2-5}
            & \multirow{2}{*}{$2250 \leq m_N$} & $\gamma$ & 400 & 1 \\
            &                       & $Z$ & $-$ & 1 \\
        \hline
    \end{tabular}
\end{table}

\begin{table}[h!]
    \caption{MuC (3 TeV) cuts (continued).}
    \label{tab:muonC3part2}
    \centering
    \begin{tabular}{|c|c|c|c|c|}
        \hline
        Process & Mass region (in GeV) & Particle ($x$) & $p_{T,x} >$ (GeV) ($<$ specified) & $|\eta|<$ \\
        \hline
        \hline

        \multirow{6}{*}{$W$-$\gamma$} 
            & \multirow{2}{*}{$m_N \leq 750$} & $\gamma$ & 200 & 1 \\
            &                       & $\ell^{\pm}$ & $-$ & 1.5 \\
        \cline{2-5}
            & \multirow{2}{*}{$750<m_N<2250$} & $\gamma$ & 350 & 1 \\
            &                       & $\ell^{\pm}$ & $-$ & 1.5 \\
        \cline{2-5}
            & \multirow{2}{*}{$2250 \leq m_N$} & $\gamma$ & 500 & 1 \\
            &                       & $\ell^{\pm}$ & $-$  & $-$ \\
        \hline
        \multirow{6}{*}{$2\ell$-$\gamma$} 
            & \multirow{2}{*}{$m_N \leq 750$} & $\gamma$ & 200 & 1.25 \\
            &                       & $\ell^{\pm}_1$ &  $500$  & $-$ \\
        \cline{2-5}
            & \multirow{2}{*}{$750<m_N<2250$} & $\gamma$ & 300 & 1 \\
            &                       & $\ell^{\pm}_1$ &  $-$  & 1.25 \\
        \cline{2-5}
            & \multirow{2}{*}{$2250 \leq m_N$} & $\gamma$ & 450 & 1 \\
            &                       & $\ell^{\pm}_1$ &  $-$  & 1.25 \\
        \hline
        \multirow{6}{*}{0-$W$} 
            & \multirow{2}{*}{$m_N \leq 750$}  & $W^{\pm}$ & 400 & $-$ \\
            &                       & $\ell^{\pm}$ & $-$ & 1.5 \\
        \cline{2-5}
            & \multirow{2}{*}{$750<m_N<2250$}  & $W^{\pm}$ & 400 & $-$ \\
            &                       & $\ell^{\pm}$ & $-$ & 1.5 \\
        \cline{2-5}
            & \multirow{2}{*}{$2250 \leq m_N$} & $W^{\pm}$ & 550 & 1 \\
            &                       & $\ell^{\pm}$ & 650  & 1.25 \\
        \hline
        \multirow{9}{*}{$\gamma$-$W$} 
            & \multirow{3}{*}{$m_N \leq 750$} & $\gamma$ & $-$ & 1 \\
            &                       & $W^{\pm}$ & 200 & 2 \\
            &                       & $\ell^{\pm}$ & 600 & 1.5 \\
        \cline{2-5}
            & \multirow{3}{*}{$750<m_N<2250$} & $\gamma$ & $-$ & 1.25 \\
            &                       & $W^{\pm}$ & 400 & 1 \\
            &                       & $\ell^{\pm}$ & 500 & 1.25 \\
        \cline{2-5}
            & \multirow{3}{*}{$2250 \leq m_N$} & $\gamma$ & $<\sqrt{s}-m_N$ & 1.25 \\
            &                       & $W^{\pm}$ & 600 & 1 \\
            &                       & $\ell^{\pm}$ &  600 & 1 \\
        \hline
        \multirow{9}{*}{$Z$-$W$} 
            & \multirow{3}{*}{$m_N \leq 750$} & $Z$ & 400 & 1 \\
            &                       & $W^{\pm}$ & 200 & $-0.5 < \eta < 1.5$ \\
            &                       & $\ell^{\pm}$ & 400 & $-1 < \eta < 1.5$ \\
        \cline{2-5}
            & \multirow{3}{*}{$750<m_N<2250$} & $Z$ & 300 & 1 \\
            &                       & $W^{\pm}$ & 400 & 1 \\
            &                       & $\ell^{\pm}$ & 400 & $1$  \\
        \cline{2-5}
            & \multirow{3}{*}{$2250 \leq m_N$} & $Z$ & $-$ & 1 \\
            &                       & $W^{\pm}$ & 800 & 1 \\
            &                       & $\ell^{\pm}$ & 800  & 1 \\
        \hline
        \multirow{9}{*}{$W$-$W$} 
            & \multirow{3}{*}{$m_N \leq 750$} & $W^{\pm}_1$ & 400 & 1 \\
            &                       & $W^{\pm}_2$ & 200 & 1 \\
            &                       & $\ell^{\pm}_1$ &  600  & 1.25 \\
        \cline{2-5}
            & \multirow{3}{*}{$750<m_N<2250$} & $W^{\pm}_1$ & 500 & 1 \\
            &                       & $W^{\pm}_2$ & 200 & 1 \\
            &                       & $\ell^{\pm}_1$ &  500  & 1 \\
        \cline{2-5}
            & \multirow{3}{*}{$2250 \leq m_N$} & $W^{\pm}_1$ & 600 & 1 \\
            &                       & $W^{\pm}_2$ & $200<p_T<\sqrt{s}-m_N$ & 1 \\
            &                       & $\ell^{\pm}_1$ &  700  & 1 \\
        \hline
        \multirow{6}{*}{$2\ell$-$W$} 
            & \multirow{2}{*}{$m_N \leq 750$}  & $W^{\pm}$ & 300 & $-$ \\
            &                       & $\ell_1^{\pm}$ & 600 & $-$ \\
        \cline{2-5}
            & \multirow{2}{*}{$750<m_N<2250$}  & $W^{\pm}$ & 400 & $-$ \\
            &                       & $\ell_1^{\pm}$ & 600 & $-$ \\
        \cline{2-5}
            & \multirow{2}{*}{$2250 \leq m_N$} & $W^{\pm}$ & 600 & $-$ \\
            &                       & $\ell_1^{\pm}$ & 650  & $-$ \\
        \hline

    \end{tabular}
\end{table}

\begin{table}[h!]
    \caption{MuC (10 TeV) cuts.}
    \label{tab:muonC10}
    \centering
    \begin{tabular}{|c|c|c|c|c|}
        \hline
        Process & Mass region (in GeV) & Particle ($x$) & $p_{T,x} >$ (GeV) ($<$ specified) & $|\eta|<$ \\
        \hline
        \hline
        \multirow{3}{*}{0-$\gamma$} 
            & \multirow{1}{*}{$m_N \leq 3000$} & $\gamma$ & 400 & $-$ \\ 
        \cline{2-5}
            & \multirow{1}{*}{$3000<m_N<7000$} & $\gamma$ & 700 & 1 \\
        \cline{2-5}
            & \multirow{1}{*}{$7000 \leq m_N$} & $\gamma$ & 1000 & 1 \\
        \hline
        \multirow{6}{*}{$\gamma$-$\gamma$} 
            & \multirow{2}{*}{$m_N \leq 3000$} & $\gamma_1$ & 400 & 1.5 \\
            &                       & $\gamma_2$ & 100 & 1 \\
        \cline{2-5}
            & \multirow{2}{*}{$3000<m_N<7000$} & $\gamma_1$ & 700 & 1 \\
            &                       & $\gamma_2$ & 100 & 1 \\
        \cline{2-5}
            & \multirow{2}{*}{$7000 \leq m_N$} & $\gamma_1$ & 1000 & 1 \\
            &                       & $\gamma_2$ & 100 & 1 \\
        \hline
        \multirow{6}{*}{$Z$-$\gamma$} 
            & \multirow{2}{*}{$m_N \leq 3000$} & $\gamma$ & 300 & $-0.5 < \eta < 1.5$ \\
            &                       & $Z$ & 700 & 1 \\
        \cline{2-5}
            & \multirow{2}{*}{$3000<m_N<7000$} & $\gamma$ & 700 & 1 \\
            &                       & $Z$ & 500 & 1 \\
        \cline{2-5}
            & \multirow{2}{*}{$7000 \leq m_N$} & $\gamma$ & 900 & 1 \\
            &                       & $Z$ & $-$ & 1 \\
        \hline
        \multirow{9}{*}{$W$-$\gamma$} 
            & \multirow{3}{*}{$m_N \leq 3000$} & $\gamma$ & 600 & 1 \\
            &                       & $W^{\pm}$ & $-$ & 1 \\
            &                       & $\ell^{\pm}$ & $-$ & 1.5 \\
        \cline{2-5}
            & \multirow{3}{*}{$3000<m_N<7000$} & $\gamma$ & 1000 & 1 \\
            &                       & $W^{\pm}$ & $-$ & 1 \\
            &                       & $\ell^{\pm}$ & $-$ & 1.5 \\
        \cline{2-5}
            & \multirow{3}{*}{$7000 \leq m_N$} & $\gamma$ & 1500 & 1 \\
            &                       & $W^{\pm}$ & $-$ & 1 \\
            &                       & $\ell^{\pm}$ & $-$  & 1.5 \\
        \hline
        \multirow{6}{*}{$2\ell$-$\gamma$} 
            & \multirow{2}{*}{$m_N \leq 3000$} & $\gamma$ & 400 & 1 \\
            &                       & $\ell^{\pm}_1$ &  $-$  & 1 \\
        \cline{2-5}
            & \multirow{2}{*}{$3000<m_N<7000$} & $\gamma$ & 800 & 1 \\
            &                       & $\ell^{\pm}_1$ &  $-$  & 1 \\
        \cline{2-5}
            & \multirow{2}{*}{$7000 \leq m_N$} & $\gamma$ & 1000 & 1 \\
            &                       & $\ell^{\pm}_1$ &  $-$  & 1 \\
        \hline
        \multirow{6}{*}{0-$W$} 
            & \multirow{2}{*}{$m_N \leq 3000$}  & $W^{\pm}$ & 500 & $-$ \\
            &                       & $\ell^{\pm}$ & $-$ & 1.5 \\
        \cline{2-5}
            & \multirow{2}{*}{$3000<m_N<7000$}  & $W^{\pm}$ & 1200 & $-$ \\
            &                       & $\ell^{\pm}$ & $-$ & 1.5 \\
        \cline{2-5}
            & \multirow{2}{*}{$7000 \leq m_N$} & $W^{\pm}$ & 1600 & 1 \\
            &                       & $\ell^{\pm}$ & 1900  & 1.25 \\
        \hline
        \multirow{6}{*}{$\gamma$-$W$} 
            & \multirow{2}{*}{$m_N \leq 3000$} 
            &                        $W^{\pm}$ & 500 & $-$ \\
            &                       & $\ell^{\pm}$ & 1500 & 1.25 \\
        \cline{2-5}
            & \multirow{2}{*}{$3000<m_N<7000$} 
            &                        $W^{\pm}$ & 1300 & $-$ \\
            &                       & $\ell^{\pm}$ & 1500 & 1.25 \\
        \cline{2-5}
            & \multirow{2}{*}{$7000 \leq m_N$} 
            &                        $W^{\pm}$ & 1800 & $-$ \\
            &                       & $\ell^{\pm}$ & 1500 & 1.25 \\
        \hline

    \end{tabular}
\end{table}

\begin{table}[h!]
    \caption{MuC (10 TeV) cuts (continued).}
    \label{tab:muonC10part2}
    \centering
    \begin{tabular}{|c|c|c|c|c|}
        \hline
        Process & Mass region (in GeV) & Particle ($x$) & $p_{T,x} >$ (GeV) ($<$ specified) & $|\eta|<$ \\
        \hline
        \hline
        \multirow{9}{*}{$Z$-$W$} 
            & \multirow{3}{*}{$m_N \leq 3000$} & $Z$ & 1000 & 1 \\
            &                       & $W^{\pm}$ & 600 & $-0.5 < \eta < 2$ \\
            &                       & $\ell^{\pm}$ & 1300 & $-1.25 < \eta < 1.5$ \\
        \cline{2-5}
            & \multirow{3}{*}{$3000<m_N<7000$} & $Z$ & 800 & 1 \\
            &                       & $W^{\pm}$ & 1100 & 1 \\
            &                       & $\ell^{\pm}$ & 1500 & $-1.25 < \eta < 1.5$  \\
        \cline{2-5}
            & \multirow{3}{*}{$7000 \leq m_N$} & $Z$ & $<\sqrt{s}-m_N$ & 1 \\
            &                       & $W^{\pm}$ & 1600 & 1 \\
            &                       & $\ell^{\pm}$ & 2000  & 1.25 \\
        \hline
        \multirow{9}{*}{$W$-$W$} 
            & \multirow{3}{*}{$m_N \leq 3000$} & $W^{\pm}_1$ & 1200 & 1 \\
            &                       & $W^{\pm}_2$ & 400 & 1 \\
            &                       & $\ell^{\pm}_1$ &  2000  & 1.25 \\
        \cline{2-5}
            & \multirow{3}{*}{$3000<m_N<7000$} & $W^{\pm}_1$ & 1400 & 1 \\
            &                       & $W^{\pm}_2$ & 500 & 1 \\
            &                       & $\ell^{\pm}_1$ &  1700  & 1 \\
        \cline{2-5}
            & \multirow{3}{*}{$7000 \leq m_N$} & $W^{\pm}_1$ & 1800 & 1 \\
            &                       & $W^{\pm}_2$ & $200<p_T<\sqrt{s}-m_N$ & 1 \\
            &                       & $\ell^{\pm}_1$ &  2300  & 1 \\
        \hline
        \multirow{6}{*}{$2\ell$-$W$} 
            & \multirow{2}{*}{$m_N \leq 3000$}  & $W^{\pm}$ & 700 & $-$ \\
            &                       & $\ell_1^{\pm}$ & 1700 & $-$ \\
        \cline{2-5}
            & \multirow{2}{*}{$3000<m_N<7000$}  & $W^{\pm}$ & 700 & $-$ \\
            &                       & $\ell_1^{\pm}$ & 1700 & $-$ \\
        \cline{2-5}
            & \multirow{2}{*}{$7000 \leq m_N$} & $W^{\pm}$ & 700 & $-$ \\
            &                       & $\ell_1^{\pm}$ & 1700  & $-$ \\
        \hline

    \end{tabular}
\end{table}

\end{document}